\documentclass[letterpaper,aps,prd,preprint,showpacs,nofootinbib,superscriptaddress]{revtex4-1}
\usepackage{color}
\usepackage{multirow}
\usepackage{float}
\usepackage{subfigure}
\usepackage{rotating}
\usepackage{amsmath,amssymb,graphicx}
\usepackage{epsf}
\usepackage{bm}

\newcommand{\nc}{\newcommand}
\nc{\postscript}[2]
{\setlength{\epsfxsize}{#2\hsize}\centerline{\epsfbox{#1}}}
\nc{\non}{\nonumber}
\nc{\hc}{\hbox {h.c.}} \nc{\re}{\hbox {Re}} \def\im{{\rm Im}}
\nc{\mev}{\hbox {MeV}} \nc{\gev}{\;\hbox {GeV}} \nc{\tev}{\;\hbox {TeV}}
\def\lsim{\mathrel{\raise.3ex\hbox{$<$\kern-.75em\lower1ex\hbox{$\sim$}}}}
\def\gsim{\mathrel{\raise.3ex\hbox{$>$\kern-.75em\lower1ex\hbox{$\sim$}}}}

\nc{\etal}{{\it et al.}}
\nc{\Lsp}{\;\;\;\;\;\;\;\;\;\;}  \nc{\LLLsp}{\lspace \lspace}
\nc{\lsp}{\;\;\;\;\;\;}
\nc{\spac}{\;\;\;}
\nc{\noi}{\noindent}
\nc{\beq}{\begin{equation}}   \nc{\eeq}{\end{equation}}
\nc{\bea}{\begin{eqnarray}}   \nc{\eea}{\end{eqnarray}}
\nc{\baa}{\begin{array}}      \nc{\eaa}{\end{array}}
\nc{\bit}{\begin{itemize}}    \nc{\eit}{\end{itemize}}
\nc{\ben}{\begin{enumerate}}  \nc{\een}{\end{enumerate}}
\nc{\bce}{\begin{center}}     \nc{\ece}{\end{center}}

\def\sq2{\sqrt{2}}

\def\ph{\varphi}

\def\m4{m^4(\ph)}
\def\mn2{m_n^2}

\def\v5{V^{(5)}}

\begin{document}

%\begin{frontmatter}

\title{\begin{flushright}  \mbox{\normalsize \rm CUMQ/HEP 189}
       \end{flushright}
       \vskip 20pt
       %Pseudoscalar Bulk Higgs as the 750 GeV diphoton signal
      Bulk Higgs and the 750 GeV diphoton signal
}
%% \author[label1]{Mariana Frank}\ead{mariana.frank@concordia.ca}
%% \author[label1,label2]{Nima Pourtolami}\ead{n\_pour@live.concordia.ca}
%% \author[label2]{Manuel Toharia}\ead{mtoharia@dawsoncollege.qc.ca}
%% \address[label1]{Department of Physics, Concordia University,
%% 7141 Sherbrooke St. West, Montreal, Quebec,\ Canada, H4B 1R6}
%% \address[label2]{Physics Department, Dawson College,
%%  3040 Sherbrooke St., Westmount, Quebec, Canada H3Z 1A4}

\author{Mariana Frank\footnote{mariana.frank@concordia.ca}}
\affiliation{Department of Physics, Concordia University\\
7141 Sherbrooke St. West, Montreal, Quebec,\\ CANADA H4B 1R6
}
\author{
Nima Pourtolami\footnote{n\_pour@live.concordia.ca}}
\affiliation{Department of Physics, Concordia University\\
7141 Sherbrooke St. West, Montreal, Quebec,\\ CANADA H4B 1R6
}
\affiliation{Physics Department, Dawson College\\
 3040 Sherbrooke St., Westmount, Quebec,\\ CANADA H3Z 1A4
}

\author{Manuel Toharia\footnote{mtoharia@dawsoncollege.qc.ca}}
\affiliation{Physics Department, Dawson College\\
 3040 Sherbrooke St., Westmount, Quebec,\\ CANADA H3Z 1A4
}

\date{\today}

\begin{abstract}
We consider scenarios of warped extra-dimensions with all matter fields in
the bulk and in which both the hierarchy and the flavor puzzles of the
Standard Model are addressed. The simplest extra dimensional
extension of the Standard Model Higgs sector, i.e a 5D bulk Higgs
doublet, can be a natural and simple explanation to the 750
GeV excess of diphotons hinted at the LHC, with  
the resonance responsible for the signal being the lightest CP odd excitation
coming from the Higgs sector. No
new matter content is invoked, the only new ingredient being the
presence of (positive) brane localized kinetic terms associated to the 5D
bulk Higgs, which allow to reduce the mass of the lightest CP odd Higgs excitation to
$750$ GeV. Production and decay of this resonance can naturally fit
the observed signal when the mass scale of the rest of
extradimensional resonances is of order $1$ TeV.

\end{abstract}

%% \begin{keyword}
%% %% keywords here, in the form: keyword \sep keyword
%% Warped Extra Dimensions, Higgs, diphoton excess 
%% %% MSC codes here, in the form: \MSC code \sep code
%% %% or \MSC[2008] code \sep code (2000 is the default)

%% \end{keyword}

%\end{frontmatter}

\maketitle

\section{Introduction}

The original motivation for warped extra-dimensions was to
address the hierarchy problem, so that the fundamental scale of gravity
is exponentially reduced along the extra dimension, from the Planck mass scale to the TeV
scale. Thus, the TeV scale becomes the natural scale of the Higgs sector
if this one is localized near the TeV boundary of the
extra dimension, as first introduced by Randall and Sundrum (RS) \cite{RS1}.
If SM fields are allowed to propagate in the extra dimension \cite{Davoudiasl:1999tf},
the scenario can also address the flavor puzzle of the SM, explaining
fermion masses and mixings from the geographical location of fields
along the extra dimension. However, processes mediated by the heavy resonances
of the 5D bulk fields, Kaluza-Klein (KK) modes,   generate dangerous
contributions to electroweak and flavor observables (including
dangerous deviations to the $Zb\bar{b}$ coupling) \cite{Burdman,RSeff,Agashe},
pushing the KK mass scale to %$\Lambda_{UV} \sim 
$5-10$ TeV \cite{Weiler}.
A popular mechanism to lower the
KK scale involves using a custodial gauge $SU(2)_R$ symmetry
\cite{custodial}, which ensures a small contribution to electroweak
precision parameters, lowering the KK scale bound to about 3 TeV.

Alternatively, one can study scenarios in which the metric is slightly
modified from the RS metric background ($AdS_5$). This can be achieved
quite naturally from the backreaction on the metric caused by a 5D scalar
field stabilizing the original $AdS_5$ warped background
\cite{Goldberger:1999uk}. When the 5D Higgs is sufficiently leaking into the
bulk and when the metric background is modified near the
TeV boundary, the scenario allows for KK scales as low as 1-2 TeV, with precision electroweak and
flavor constraints under control \cite{Cabrer}.
An inconvenience is that these scenarios are typically hard to probe
experimentally as the couplings of all particles are very suppressed \cite{Falkowski:2008fz,deBlas:2012qf}. 
Still, it has been shown that it can
still lead to interesting deviations in Higgs phenomenology, as the Higgs couplings
can receive sufficient radiative corrections from the many KK fermions
of the model \cite{Frank:2015zwd}. 

Run 2 LHC data at $\sqrt{s}=13$ TeV shows signals of a new resonance in
the diphoton distribution at an invariant mass of 750 GeV with a 3.9
significance at ATLAS \cite{ATLAS750}, with 3.2 fb$^{-1}$ and 3.4
combined significance at CMS \cite{CMS:2016owr},  
with 2.6
fb$^{-1}$(combining run 1 and run 2 results). ATLAS reports 14 events
and CMS, 10. The experimental data is summarized in Table \ref{tab:siglims}.
\begin{table}
\begin{center}
\begin{tabular}{|c|c|c|}
\hline\hline
Channel & $8~{\rm TeV}, \sigma_{max}$ & $13~{\rm TeV}, \sigma_{max}$\\
  \hline\hline 
 $\gamma\gamma$   &$.21\pm.22~{\rm fb}$ \cite{ATLAS750} & $5.5\pm 1.5~{\rm fb}$  \cite{ATLAS750} \\
   &  $.63\pm .31~{\rm fb}$ \cite{CMS:2016owr} &$4.8\pm 2.1~{\rm fb}$ \cite{CMS:2016owr}  \\
   \hline \hline
% $WW$ & $46~fb$ \cite{Khachatryan:2015cwa} & $300~fb$ \cite{ATLAS-CONF-2016-021}\\
% \hline
% $ZZ$ & $12~fb$ \cite{Aad:2015kna} & $120~fb$ \cite{ATLAS-CONF-2016-016}\\
 \hline
 $t {\overline t}$ & $700~{\rm fb}$ \cite{Aad:2015fna} &   $\sim 2000~{\rm fb}$ \\
% \hline 
% $Z\gamma $ &$3.5~fb$ \cite{Aad:2014fha} & $30~fb$ \cite{ATLAS-CONF-2016-010} \\
% \hline
% $\tau^+\tau^-$ & $14~fb$ \cite{Aad:2014vgg} &$60~fb$ \cite{ATLAS-CONF-2015-061} \\
 \hline
 $jj$ & $2.1~{\rm pb}$ \cite{Khachatryan:2016ecr}& $\sim 10~{\rm pb}$  \\ 
% \hline
 %$h_1h_1$ & $41~fb$ \cite{Khachatryan:2015yea}& $154~fb$ \cite{ATLAS-CONF-2016-017}\\
 \hline
 $hZ$ &  $19~{\rm fb}$ \cite{Aad:2015wra} & $\sim 100~{\rm fb}$ \\
 \hline\hline
  \end{tabular}
 \caption{\label{tab:siglims} LHC diphoton cross sections at  ATLAS
   and CMS  and relevant signal limits for the $750$ GeV CP odd scalar
   resonance considered here. At 13 TeV, these are the rough implied bounds from the 8 TeV limits.} \end{center}
 \end{table}

%The
%anomalous events are not accompanied by significant missing energy,
%nor leptons or jets, and no resonances at invariant mass 750 GeV are
%seen in $ZZ, l^+l^-$, or $jj$.  A confirmation of the 750 GeV excess
%could be a strong indicator of new strong interactions at the TeV
%scale.  

In light of all this,
%% the recent observation by both ATLAS and CMS of an excess
%% in the diphoton channel at 750 GeV
we propose here a simple and
economic explanation within warped extra-dimensional models.
It would require the presence of a 5D bulk Higgs, and because the mass
of the new resonance is 750 GeV, the Higgs should be as much delocalized as
possible from the TeV brane (but still close enough to address the
hierarchy problem). The reason is that the masses of the Higgs KK
excitations will increase as the Higgs is pushed towards the brane,
getting infinitely heavy in the limit of a brane Higgs. 
Out of these excitations some are CP odd scalars, making them a
natural candidate for the signal since they do not couple at
tree-level to $ZZ$ or $WW$.
We will show that if the typical mass of the KK gluon (typically the lightest and most
visible KK particle) is around 1-2 TeV, it is very simple to
obtain a 750 GeV CP odd Higgs with the help of small (and positive)
brane localized kinetic terms of the 5D Higgs.  
Since the CP odd scalars do not couple at tree-level to $ZZ$ or $WW$, the largest
coupling is going to be to pairs of tops. As will be shown, this coupling can be
naturally small in wide regions of the allowed parameter space. This way, the radiative coupling to gluons, large enough
for producing CP odd scalars, can also dominate the decays and the
(also) radiative decay into photons can then receive enough branching fraction. 

%% Explanations within a warped scenario have appeared previously. The
%% 750 GeV resonance has been interpreted as a radion
%% \cite{Ahmed:2015uqt}, (and/or dilaton \cite{Cao:2016udb}), as a KK
%% graviton with positive \cite{Falkowski:2016glr} or negative
%% \cite{Hewett:2016omf} brane kinetic terms, as a new spin-2 resonance
%% \cite{Carmona:2016jhr}, or a 5D field-related axion
%% \cite{Chakrabarty:2016hxi}. The  most robust explanation in terms of
%% agreement with all experimental data, and which  preserves all the
%% attractive features of the warped space models is introducing an
%% additional scalar into the model \cite{Bauer:2016lbe}, with the
%% drawback that the model is not minimal anymore. We show below that out
%% explanation, while preserving minimality and agreement with the
%% diphoton excess and Higgs data, is satisfied in a significant region
%% of the parameter space, rendering it natural. 
Explanations of the 750 GeV signal within warped scenarios have
been put forward previously, with the resonance interpreted as a radion
\cite{Ahmed:2015uqt}, (and/or dilaton \cite{Cao:2016udb}), as a KK 
graviton \cite{Falkowski:2016glr,Hewett:2016omf,Carmona:2016jhr}, a 5D
field-related axion \cite{Chakrabarty:2016hxi} or as an additional
5D singlet scalar added to the model \cite{Bauer:2016lbe}.
The explanation proposed here, while preserving
minimality and agreement with the diphoton excess, is also satisfied
naturally in a significant region of the parameter space.

We proceed as follows. In Sec. \ref{sec:background} we describe briefly the warped scenario, followed by its Higgs and gauge sector in Sec. \ref{sec:higgsgauge}, and of the CP-odd sector in more detail in Sec \ref{sec:cpodd}. Within that section we look at the fermion couplings in \ref{subsec:fermion}, the $\gamma \gamma$ and $glu-glu$ couplings in \ref{subsec:gggamgam} and to $Zh$ couplings in \ref{subsec:zh}. Our numerical estimates are presented in \ref{subsec:numerical} and we conclude in Sec. \ref{sec:discussion}. We leave some of the details for the Appendix.

\section{The background metric}
\label{sec:background}
The (stable) static spacetime background is:
\bea
ds^2 = e^{-2\sigma(y)}\eta_{\mu\nu} dx^\mu dx^\nu - dy^2,
\label{RS}
\eea
where the extra coordinate $y$ ranges between the
two boundaries at $y=0$ and $y=y_{1}$, and where $\sigma(y)$ is the warp factor
responsible for exponentially suppressing mass scales at different
slices of the extra dimension. In the original RS scenario, $\sigma(y)=ky$,
with $k$ the curvature scale of the $AdS_5$ 
interval that we  take of the same order as $M_{Pl}$. Nevertheless
this configuration is not stable as it contains 
a massless radion, a result of having the length of the interval not
fixed. In more general warped scenarios with stabilization mechanism,
$\sigma(y)$ is a more general (growing) function of $y$.

We  consider here the specific case where a 5D bulk stabilizer field
backreacts on the $AdS_5$ metric producing the warp factor \cite{Cabrer,Carmona}. 
\bea\label{genericmetric}
\sigma(y) = ky - {1\over \nu^2} \text{log}\left(1-{y\over y_s}\right), 
%\phi(y) = -{\sqrt{6}\over \nu} \text{log}\left[\nu^2 bk(y_s - y)\right],
\eea
where $y = y_s$ is the position of a metric singularity, which stays
beyond the physical interval considered here, i.e. $y_s> y_{1}$.
%% For this geometry the curvature $kL$ and the
%% curvature radius $R$ are modified from the $AdS_5$ case 
%% and are given by
%% \bea\label{curvature}
%% kL(y) = { k \Delta \nu^2\over\sqrt{1-2\nu^2/5 + 2k\Delta\nu^2+(k\Delta)^2\nu^4}},
%% \eea
%% and
%% \bea
%% R(y) = -20 k^2{ (1 - 2/5 \nu^2 + 2 k\Delta\nu^2 +  (k\Delta)^2\nu^4\over (k\Delta)^2\nu^4},
%% \eea
%% respectively, where $k\Delta = k(y_s - y)$ is always positive, \ie,~ the singularity is always assumed to
%% be outside of the physical region. We refer to these scenarios as
%% the modified $AdS_5$ ($MAdS_5$) scenarios. 
%% %
In these modified metric scenarios, the Planck-TeV hierarchy is reproduced with a
shorter extra-dimensional length due to a stronger warping near the
TeV boundary, so that whereas in RS we have $k y_{1}\simeq 35$, in
the modified scenarios we can have $k y_{1}\simeq 20-30$. The appeal
of this particular modification lies on the possibility of allowing
for light KK particles ($\sim 1$ TeV), while 
keeping flavor and precision electroweak bounds at bay. This happens
when the Higgs profile leaks sufficiently out of the TeV brane so
that all of its couplings to KK particles are suppressed compared to
the usual RS scenario \cite{Cabrer, Falkowski:2008fz,Carmona}. 
We  thus fix the Higgs localization to a point where it is
maximally pushed away from the IR brane, while still solving the
hierarchy problem (i.e. making sure that we are not reintroducing a
new fine-tuning of parameters within the Higgs potential parameters \cite{Cabrer,Quiros:2013yaa}.)

\section{Gauge and Higgs sector}
\label{sec:higgsgauge}
The matter content of the model is that of a minimal 5D extension of the
Standard Model, so that we assume the usual strong and
electroweak gauge groups $SU(3)_c\times SU(2)_L\times U(1)_Y$, with all
fields propagating in the bulk. The fermions of 
the model are also bulk fields, with different 5D bulk masses, so that
their zero mode wavefunctions are localized at different sides of the
interval. This way the scenario also addresses the flavor puzzle of
the SM, since hierarchical masses and small mixing angles for the 
SM fermions become a generic feature due to fermion localization and
small wavefunction overlaps \cite{Frank:2015sua}.

In the electroweak and Higgs sector we consider the following action
\bea
\hspace{-1cm}&&S=\int d^4xdy \sqrt{g}\ \left(-\frac{1}{4} F^2_{MN} +
|D^M H|^2 - V(H) \right) \ \ \ \ \\
\hspace{-1cm} && + \sum_{i=1}^{2}\int d^4xdy \sqrt{g}\ \delta(y-y_i) \left(\frac{d_i}{k}  |D^M H|^2 - \lambda_i(H)\right)   
\label{5Daction}
\eea
where the capital index $M$ will be used to denote the $5$ spacetime directions, while the
Greek index $\mu$ will be used for the 4D directions. The
coefficients $d_i$ (in units of $k$) are essentially free parameters
encoding the importance of brane localized kinetic terms associated with the bulk  
Higgs field. These terms will allow for a slight modification of the spectrum of the KK Higgs
excitations, particularly useful in reproducing a $750$ GeV CP-odd
excitation. These brane kinetic terms can be thought of as exactly
localized operators, or as bulk operators that happen to be
dynamically localized due to couplings to some localizer
VEV\footnote{In order to avoid tachyons and/or ghosts,
  the sign of the purely brane localized brane kinetic terms will be
  kept positive, i.e. $d_i>0$.}.

The 5D Higgs doublet is expanded around a nontrivial VEV profile
$v(y)$ as
\bea
\label{Hexpansion}
H=\frac{1}{\sqrt{2}} e^{i g_5 \Pi}\left(\begin{matrix} 0\\ v(y) +h(x,y) \end{matrix}
\right)
\eea
and the covariant derivative is $D_M= \partial_M + i g_5 A_M$ with 
\bea
\label{Aexpansion}
A_M=\left(
\begin{matrix}
  s_W A_M^{em}+\frac{c_W^2-s_W^2}{2c_W}Z_M & \frac{1}{\sqrt{2}}W_M^+\\
 \frac{1}{\sqrt{2}}W_M^-& -\frac{1}{2c_W} Z_M  
 \end{matrix}
\right)
\eea
and CP-odd and charged Higgs part is  
\bea
\label{Piexpansion}
\Pi=\left(
\begin{matrix}
\frac{c_W^2-s_W^2}{2c_W}\Pi_z & \frac{1}{\sqrt{2}} \Pi^+\\
 \frac{1}{\sqrt{2}}\Pi^-& -\frac{1}{2c_W} \Pi_z    
\end{matrix}
\right)
\eea
with the weak angle defined like in the SM, i.e. $s_W/c_W=g'_5/g_5$,
where $g_5$ and $g_5'$ are the 5D coupling constants of $SU(2)_L$ and $U(1)_Y$.

The extraction of degrees of freedom in this context has been
performed in \cite{Quiros:2013yaa,Falkowski:2008fz,Archer:2012qa} and we outline
here the main results.  The effect of brane kinetic terms in the
Higgs sector is new and its derivation is outlined in the Appendix.
The 5D equations of motion for all these fields are coupled
(except for the case of the real Higgs excitation $h(x,y)$) and in order to
decouple them, one can partially fix the gauge, or add a gauge fixing term to the
previous 5D action.
For example, in the CP-odd case, the fields $Z_\mu(x,y)$,
$Z_5(x,y)$ and $\Pi_z(x,y)$ must be unmixed.
The partial gauge fixing constraint\footnote{There is still be some
  gauge freedom left, so that the towers of 4D Goldstone bosons that 
  appear can be gauged away.}
\bea
\partial^\mu Z_\mu - M^2_z(y) \Pi_z +(e^{-2\sigma}Z_5)'=0
\eea
manages to decouple the fields $Z_\mu$ from $Z_5$ and
$\Pi_z$ in the bulk. We defined here  $M_z(y)=\frac{g_5}{2c_W} v(y) e^{-\sigma(y)}$.

However, the presence of the Higgs brane kinetic terms,
proportional to $d_i$ in the action,  forces us to extend the gauge
choice on the branes, producing a lifting of $Z_5$ field so that the
decoupling is maintained at the boundaries\footnote{In the
absence of brane kinetic terms, $Z_5$ must have vanishing boundary
conditions (Dirichlet) if $Z_\mu$ is to have Neumann conditions and
thus develop a zero mode KK excitation in the effective 4D theory.}.
The appropriate boundary condition at the IR brane is
\bea
Z_5(x,y_1) =- \frac{d_1}{k} M^2_z(y_1) e^{2\sigma(y_1)} \Pi_z(x,y_1)
\eea
where $y_1$ denotes the position of the boundary (note that if the
brane kinetic term parameter $d_1$ tends to zero, the condition on $Z_5$
becomes Dirichlet, as expected).
With this type of gauge choice, the 5D fields $Z_\mu$, $W_\mu$ and
$A_\mu$ have independent 5D equations of motion. In order to extract
the effective 4D degrees of freedom, we expand the gauge fields as
$Z_\mu(x,y)=Z^n_\mu(x) f^n_z(y)$, $W_\mu(x,y)=W^n_\mu(x) f^n_w(y)$ 
and $A_\mu(x,y)=A^n_\mu(x) f^n_\gamma(y)$ (summation over $n$ is understood)
and where $Z^0_\mu(x)$, $W^0_\mu(x)$ and $A^0_\mu(x)$ are 
the $Z$, $W$ and $\gamma$ gauge bosons of the SM.
The extradimensional profiles $f^n_z(y)$, $f^n_w(y)$ and $f^n_\gamma(y)$
are solutions of
\bea\label{gaugeeq}
\left(e^{-2\sigma} f'_a\right)'+(m^2_{n} - M_a^2(y)) f_a =0
\eea
where $a=z,w,\gamma$ and where $M_z(y)=\frac{g_5}{2c_W} v(y) e^{-\sigma(y)}$, as defined before, 
$M_w(y)=\frac{g_5}{2} v(y) e^{-\sigma(y)}$,  and  $M_\gamma=0$.
The boundary conditions for these profiles are\footnote{We ignore here
  possible brane localized gauge kinetic terms and keep only the
  effects from Higgs brane kinetic terms. We include everything in the
  derivation outlined in the Appendix.} 
\bea\label{gaugebc}
\frac{d_i}{k} M_a^2 f_a(y_i) = - e^{-2\sigma}f'_a(y_i).
\eea
The CP-even Higgs field is expanded as $h(x,y)=h^n(x) h^n_y(y)$ and the
equations for the Higgs profiles are, with $h_y \equiv h_y^n(y)$:
\bea
\left(e^{-4\sigma} h'_y\right)'+(m^2_{h_n} - \mu^2_{bulk}) h_y =0, 
\eea
where $\displaystyle \mu^2_{bulk} = \frac{\partial^2 V }{\partial H^2}\Big|_{H=v}$.
The boundary conditions are
\bea
\left(\mu^2_{brane_i} - \frac{d_i}{k}\ m^2_{h_n} e^{2\sigma}\right)  h_y =
-e^{-4\sigma} h'_y, 
\eea
with  $\displaystyle \mu^2_{brane_i} = \frac{\partial^2 \lambda_i
}{\partial H^2}\Big|_{H=v}$. Note that the mode $h^0(x)$ is 
interpreted as the SM Higgs boson.

There are still some degrees of freedom left, and their 5D equations of
motion still happen to be mixed. One of the coupled systems involves $Z_5$ and
$\Pi_z$ and the other coupled system involves $\Pi^{\pm}$ and
$W_5^\pm$.
In order to disentangle these systems one must perform a mixed
expansion, so that the decoupling of fields will happen KK level by KK
level.
The mixed expansions are, in the CP-odd sector,
\bea\label{Z5decomp}
\hspace{-.6cm}Z_5(x,y)&=& G^n(x)\frac{f'_{G_n}(y)}{m^2_{G_n}} +
\Pi_n(x)\frac{e^{2\sigma}}{m^2_{\pi_n}}X_\pi(y)   \\ 
\hspace{-.6cm}\Pi_z(x,y)&=& G^n(x)\frac{f_{G_n}(y)}{m^2_{G_n}} + \Pi_n(x)
\frac{1}{m^2_{\pi_n}M^2_{z} }X_{\pi}'(y),
\label{Pidecomp}
\eea
and in the charged scalar sector they are
\bea
\hspace{-.6cm}W_5^\pm(x,y) &=& G_n^\pm(x) \frac{f_{G_n^\pm}(y)}{m^2_{G_n^\pm}} +
\Pi^\pm_n(x)\frac{e^{2\sigma}}{m^2_{\pi_n^\pm}}X_\pm(y)\ \  \\
\hspace{-.6cm}\Pi^\pm(x,y)&=& G_n^\pm(x)  \frac{f_{G_n^\pm}(y)}{m^2_{G_n^\pm}}
+\Pi^\pm_n(x) \frac{1}{m^2_{\pi_n^\pm}M^2_{w} }X_\pm'(y).\ \ 
\eea
where  $M_z(y)$ and $M_w(y)$ were defined  below Eq~(\ref{gaugeeq}).

The effective 4D physical fields are the tower of CP-odd neutral scalars
$\Pi_n(x)$ and the tower of charged scalars $\Pi^\pm_n(x)$.
Their associated extra-dimensional profiles $X_\pi(y)$ and $X_\pm(y)$
obey the equations
\bea
\left(\frac{1}{M_a^2(y)} X_a' \right)'+ \left(\frac{m^2_{\pi_a}}{M_a^2(y)}- 1 \right) e^{2\sigma} X_a=0
\label{Xeq}
\eea
where $M_a(y)=(M_z(y),M_{w}(y))$ and $X_a=(X_\pi,X_{\pm})$. The boundary conditions are
\bea\label{Xbc}
\frac{d_i}{k}\ X'_a= -X_a,
\eea
and note that vanishing Higgs brane kinetic terms implies Dirichlet
boundary conditions for $X_a$. We  checked that these bulk
equations agree with  \cite{Falkowski:2008fz,Quiros:2013yaa,Archer:2012qa}, the
only new addition being the boundary conditions imposed by the presence of Higgs
brane kinetic terms. 

In order for these 4D scalars to be canonically normalized, we require
\bea
\frac{1}{m^2_a}\int dy\ e^{2\sigma} \frac{X_a^2}{M_a^2}=1
\eea
and this condition includes the effect of Higgs brane kinetic terms.

The remaining 4D fields are $G_n(x)$ and $G_n^\pm(x)$, which are
Goldstone bosons at each KK level. The profile wavefunctions 
$f_{G_a}(y)$ obey the same differential equations as the gauge
profiles, Eq.(\ref{gaugeeq}), as well as the same boundary
conditions, Eq.(\ref{gaugebc}). The spectrum is thus identical
to the gauge bosons spectrum level by level. These fields  appear in the
effective 4D action coupled to ($\partial^\mu Z^n_\mu$) or ($\partial^\mu W^n_\mu$), and of course
there is a leftover gauge freedom allowing us to gauge them
away (i.e., they are pure gauge).  

We wish to identify the lightest CP-odd scalar $\Pi_0(x)$ with the
observed diphoton peak at the LHC, so that we need to fix its mass $m_{\Pi_0}=  
750$~GeV. In order to have an idea of the effects of the Higgs brane kinetic terms on the
CP-odd scalar spectrum, we consider two different parameter points,
one in which the RS background metric is recovered with $\nu=10$ and 
$y_s=4\times y_{1}$ ($\nu$ is the exponent appearing in the modified
metric and if relatively large, the location of spurious
singularity is sent away from the boundary, recovering essentially the
$AdS_5$ metric). The other case is 
the situation where the metric modification allows for TeV size KK
masses, which are safe from precision electroweak constrains. The
parameters chosen there are  $\nu=0.5$ and $y_s=1.04\times y_{1}$. 
In both parameter points, we fix the KK mass of the first gluon
excitation to be $1500$ GeV\footnote{Of course, this RS point is presented for comparison
only, since such light KK masses will produce too large deviations  in the
precision electroweak observables.}. 

\begin{figure}[t]
\center
\begin{center}
  \includegraphics[height=9.5cm]{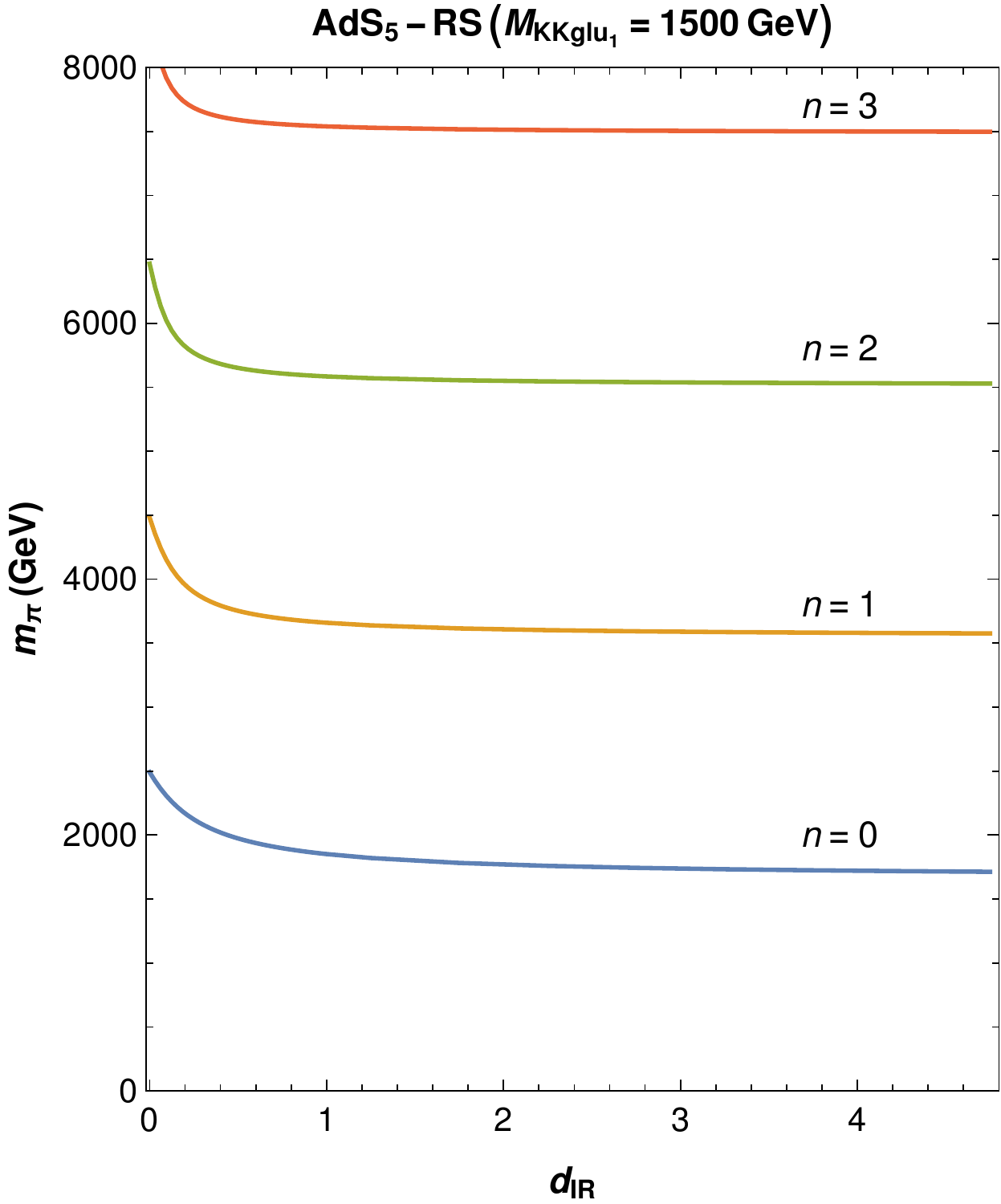}\ \ 
  \includegraphics[height=9.5cm]{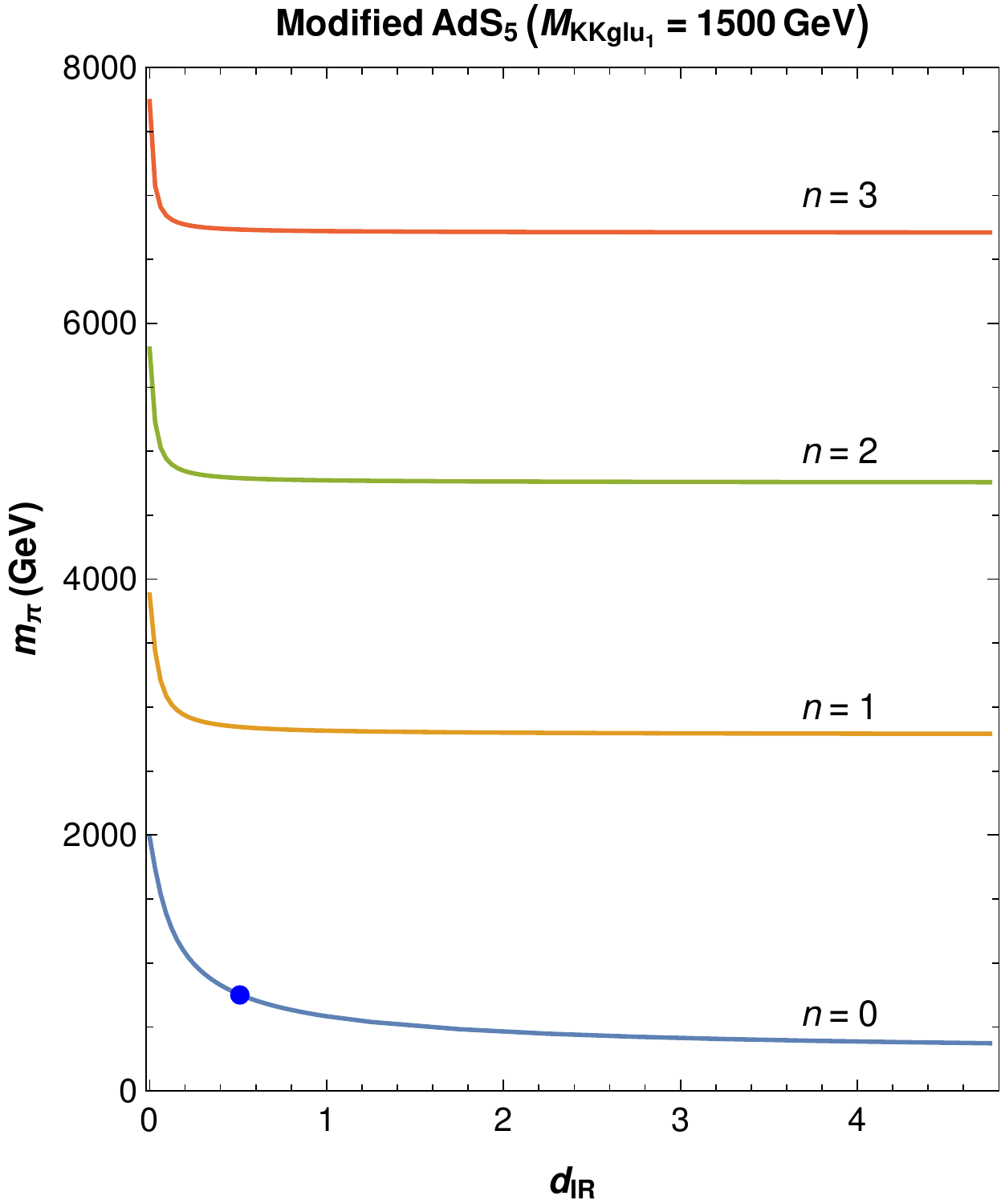}
\end{center}
\vspace{-.2cm}
\caption{Mass spectrum of the first CP-odd Higgs scalars as a function
  of the brane kinetic term coefficient $d_{1}$ in the RS metric limit
  ($\nu=10$ and $y_s=4\ y_{1}$; left panel) and within a noticeably
  IR-modified $AdS_5$ metric ($\nu=0.5$ and $y_s=1.04\ y_{1}$; right
  panel). In both cases, the first KK gluon mass  is fixed at 1500
  GeV. One can see that in RS it is not possible to obtain a 750 GeV CP-odd scalar mass,
  whereas within the modified metric, a brane kinetic coefficient
  $d_{1}\simeq 0.51$ does produce a mass  of 750 GeV, shown here as a dot. }
\label{fig:Xmassplot}
\vspace{-.2cm}
\end{figure}

In Fig. \ref{fig:Xmassplot} we show the spectrum of the first 4 KK levels of CP-odd Higgs bosons 
$\Pi_n(x)$, for $n=0,1,2,3$, as a function of the brane kinetic term
$d_{1}$ in units of the curvature $k\sim M_{Pl}$. The effects of the
UV localized brane kinetic term are warped suppressed and so we do not
consider them here anymore.
We can see that in the RS limit, it is not possible to reduce
sufficiently the lightest CP-odd mass to 750 GeV, as its mass tends 
asymptotically from about $2500$ GeV without brane kinetic term to about
$1750$ GeV for large brane kinetic terms.
On the other hand, with the modified metric it becomes possible to
reduce greatly the lightest CP-odd mass 
with relatively small brane kinetic term coefficients (which in this
particular case tends asymthotically to about 400 GeV).
Parameter points such that the metric modification lies between the
two considered will actually have an intermediate behavior, with a
lightest CP-odd mass having  increasing asymptotic values as one
recovers the RS background. 

Finally also note that the
  spectrum for the charged scalars is essentially the same since their
  differential equations and boundary conditions are identical except for the
  functions $M_z(y)$ and $M_w(y)$, which differ by about $10\%$. $M_w(y)$
  produces a deviation from the CP-odd scalar spectrum of less than
  $5\%$. This means that the scenario under consideration should also contain
  a lightest charged Higgs scalar with a mass of about $750$ GeV also.
  
The next question to ask is how big is the effect of the Higgs brane
kinetic term on the gauge bosons, and in particular on the lowest
ones, i.e. the SM $W$ and $Z$ bosons. These terms represent an
additional (brane localized) contribution to the mass of the gauge
bosons. In principle their mass is generated here from a bulk Higgs
mechanism, unless the brane kinetic terms are overly important (not
the limit we are working with here). We can quickly estimate  
its effect on the lowest lying gauge fields. These are essentially
flat (like all gauge zero modes) and thus their wave function is
$f^0_z\sim 1/\sqrt{y_{1}}$. The contribution of a brane localized
mass squared term is  $\delta m^2_z \lsim d_{1}\
v^2/y_{1}\sim d_{1}\times 700$ GeV$^2 $. For IR brane kinetic term
coefficient $d_{1}$ of ${\cal O}(1)$, this represents naively at
most some  $10\% $ contribution to the overall mass squared of either $W$ or
$Z$.  In the particular case of the
modified metric with a brane coefficient $d_{1}\simeq 0.51$, and
metric parameters $\nu =0.5$ and $y_s=1.04 \times y_1$ (which
produces a light CP-odd scalar of $750$ GeV), the exact numerical
effect on the zero mode gauge boson masses ($W$ and $Z$) is a shift of
$3$ GeV with respect to the no-brane-kinetic-term limit. Of course, in the
presence of brane kinetic terms, one redefines 
the VEV normalization constant, and the value of $g_5$, in order to correctly
account for the SM gauge boson masses and electroweak couplings.

\section{CP-odd Higgs couplings}
\label{sec:cpodd}

As a $750$ GeV CP-odd Higgs scalar $\Pi_0(x)$ is allowed  in the spectrum, thanks
to the effect of small brane localized Higgs kinetic terms, we 
now study its couplings to SM particles in order to see if the observed excess
at the LHC can be associated with this excitation.
Of course being a CP-odd scalar its tree-level couplings to $ZZ$ and
$WW$ are zero, making it an ideal candidate for the observed exotic
events. 
We thus need to focus on its tree-level couplings to fermions (and top quark in
particular), to $Zh$ (where $h$ is the $125$ GeV Higgs) and to its
radiative couplings to photons and gluons. We study these in the subsequent subsections.

\subsection{Fermion couplings}
\label{subsec:fermion}

The couplings of  $\Pi_0(x)$ to fermions arise from two sources in
the action. First source comes from the 5D Higgs Yukawa couplings, and second, from
the gauge fermion couplings. This is because the physical field
$\Pi_0(x)$ contains some of CP-odd Higgs scalar, and some
 of $Z_5$ excitation, where $Z_5$ is the fifth component of the 5D
vector boson $Z_M$.
However the 5D Yukawa coupling allows for direct
coupling of $\Pi_0(x)$ to two zero-mode fermions, whereas the
gauge-fermion coupling allows only couplings between fermion zero-modes
and higher KK fermion levels. As we will see, it is important to keep
both couplings, since after electroweak symmetry breaking the
physical SM fermions (top quarks in particular) are mostly
zero-modes but also contain a small amount of higher KK excitations, and
could thus inherit some of the original gauge-fermion coupling, especially if
the tree-level Yukawa coupling between $\Pi_0(x)$ and zero-mode top quarks
is suppressed (as it can be).

The relevant terms in the action are the 5D Higgs Yukawa couplings and
the fermion gauge interaction term,
\bea
S_{ff\Pi_0} \subset \int d^4xdy \sqrt{g}\left[
Y_u HQU + Y_d HQD + h.c.+ \bar{Q} {\cal D}\hspace{-3mm}/\  Q + \bar{U}
{\cal D}\hspace{-3mm}/\  U +   \bar{D} {\cal D}\hspace{-3mm}/\ D \right]
\eea
where $Q,U,D$ represent the 5D fermion $SU(2)_L$ doublets,  up-type
and  down-type singlets (with generation indices and isospin indices suppressed).
The kinetic terms contain the 5D covariant derivative and from them we extract
 the terms containing the CP-odd component $Z_5(x,y)$, and from the
Higgs Yukawa couplings  we extract the terms containing the CP-odd Higgs
component $\Pi_z(x,y)$.
%\bea
%S_{ff\Pi_0} \subset \int d^4xdy \sqrt{g}\left[
%Y_u HQU + Y_d HQD + h.c.+ \bar{Q} {\cal D}\hspace{-3mm}/\  Q + \bar{U}
%{\cal D}\hspace{-3mm}/\  U +   \bar{D} {\cal D}\hspace{-3mm}/\ D \right]
%\eea

We follow the approach of \cite{Frank:2015zwd,Diaz-Furlong:2016ril} and compute these
couplings by considering only the effects of three full KK levels,
i.e. computing $21 \times 21$ fermion Yukawa coupling matrices (with 3 $up$
and 3 $down$ families, each containing zero modes and 3 KK levels with an $SU(2)_L$
doublet and a singlet in each level, i.e., 3 zero modes plus $3\times 3
\times 2$ KK modes). Note that we are interested in the couplings
of the $750$ GeV CP-odd scalar $\Pi_0(x)$ to SM fermions (top quarks
primarily), but we also need its couplings with the rest of KK
fermions, since these interactions will be crucial to generate large
enough radiative couplings to photons and gluons.

We first write the effective 4D up-type quark mass matrix as
\bea
\left(\begin{matrix}
  q^0_L(x)\ Q_L(x)\ U_L(x)
\end{matrix}\right)\  M_u\ \left(\begin{matrix}
  u^0_R(x)\\ Q_R(x)\\ U_R(x)
\end{matrix}\right) 
\eea
in a basis where $q^0_L(x)$ and $u_R^0(x)$ represent three zero-mode flavors
each (doublets and singlets of $SU(2)_L$), and  $Q_L(x)$ and $Q_R(x)$ represent three
flavors and three KK levels of the vector-like KK up-type doublets,
and $U_L(x)$ and $U_R(x)$ represent three flavors and three KK levels of
vector-like KK up-type singlets. 
The mass matrix is thus
\bea
M_u= \left(\begin{matrix}
  (y^{0}_{u})_{3\times3}      &  (0)_{3\times 9}   & (Y^{qU})_{3\times 9}\\
 (Y^{Qu})_{9 \times3}      & (M_Q)_{9\times  9} & (Y_1)_{9 \times 9}\\
(0)_{9 \times3}      &     (Y_2)_{9\times  9} &  (M_U)_{9\times 9}
\end{matrix}\right)
  \label{mumatrix}
\eea
with the down sector mass matrix ${M}_d$ computed in the same way.

The submatrices are obtained by evaluating the overlap integrals 
\bea\label{YYY1KK}
y^0_u= \frac{(Y^{5D}_u)_{ij}}{\sqrt{k}} \int_0^{y_1} dy e^{-4\sigma(y)}
\frac{v(y)}{\sqrt{2}} q^{0,i}_{L}(y)u^{0,j}_{R}(y) \\ 
Y^{qU}= \frac{(Y^{5D}_u)_{ij}}{\sqrt{k}} \int_0^{y_1} dy e^{-4\sigma(y)}
\frac{v(y)}{\sqrt{2}} q^{0,i}_{L}(y)U^{n,j}_{R}(y) \\ 
Y^{Qu} = \frac{(Y^{5D}_u)_{ij}}{\sqrt{k}} \int_0^{y_1} dy e^{-4\sigma(y)}
\frac{v(y)}{\sqrt{2}} Q^{m,i}_{L}(y)u^{0,j}_{R}(y)\\ 
Y_1 = \frac{(Y^{5D}_u)_{ij}}{\sqrt{k}} \int_0^{y_1} dy e^{-4\sigma(y)}
\frac{ v(y)}{\sqrt{2}} Q^{m,i}_{L}(y)U^{n,j}_{R}(y) \\ 
Y_2 = \frac{(Y^{5D^*}_u)_{ij}}{\sqrt{k}} \int_0^{y_1} dy e^{-4\sigma(y)} \frac{ v(y)}{\sqrt{2}} Q^{m,i}_{R}(y)U^{n,j}_{L}(y)\, ,
\eea
where the indices $m$ and $n$ track the KK level and $i,j=1,2,3$ are
5D flavor indices.
The diagonal matrices  $(M_Q)_{9\times  9}$ and  $(M_U)_{9\times 9}$
are constructed with the masses of all the KK quarks involved. The
masses and the profiles of the KK fermions appearing in these overlap
integrals ($Q_L(y)$, $Q_R(y)$, $U_L(y)$ and $U_R(y)$) are obtained
by solving  differential equations for the fermion profiles
\bea
\partial_y\left(e^{(2c-1)\sigma(y)}\partial_y\left(
e^{-(c+2)\sigma(y)}\right)\right)f(y)+e^{(c-1)\sigma(y)} m_n^2 f(y) =0
\label{fermionprofile}
\eea
where $f(y)$ is the KK profile. The mass eigenvalues $m_n$ are found by
imposing Dirichlet boundary conditions on the wrong chirality modes.  

As mentioned before, we have included 3 full KK levels so that the
mass matrices in the gauge basis are $21 \times 21$ dimensional matrices,
which are not diagonal.
One needs to diagonalize them, and by doing so, move to the quark
physical basis where all the fermions couplings can then be extracted.

In the CP-odd scalar sector, we can write the effective 4D Yukawa-type couplings to fermions
in the same gauge basis as before
\bea
\left(\begin{matrix}
  q^0_L(x)\ Q_L(x)\ U_L(x)
\end{matrix}\right)  {\bf Y}_\pi \left(\begin{matrix}
  u^0_R(x)\\ Q_R(x)\\ U_R(x)
\end{matrix}\right) \ \Pi_0(x)
\eea
where now the $21 \times 21$ coupling matrix ${\bf Y}_{\pi}$ is given by 
\bea
{\bf Y}_{\pi}= \left(\begin{matrix}
  (y^{0}_{\pi qu})_{3\times3}      &  (a_{\pi qQ})_{3\times 9}   & (Y_{\pi qU})_{3\times 9}\\
 (Y_{\pi Qu})_{9 \times3}      & (a_{\pi QQ})_{9\times  9} & (Y_{1}^{\pi})_{9 \times 9}\\
(a_{\pi uU})_{9 \times3}      &     (Y_{2}^{\pi})_{9\times  9} &   (a_{\pi UU})_{9\times 9}
\end{matrix}\right)
  \label{Ypimatrix}
\eea
The submatrices are obtained by the overlap integrals 
\bea\label{YYPI}
y^{0}_{\pi qu} &=&  i \frac{(Y^{5D}_u)_{ij}}{\sqrt{2k}} \int_0^{y_1} dy
e^{-3\sigma(y)} q^{0,i}_{L}(y)u^{0,j}_{R}(y) \frac{X_\pi'(y)}{m^2_{\pi_0} M_z(y)}   \\
Y_{\pi qU}&=& i\frac{(Y^{5D}_u)_{ij}}{\sqrt{2k}} \int_0^{y_1} dy
e^{-3\sigma(y)} q^{0,i}_{L}(y)U^{n,j}_{R}(y)\frac{X_\pi'(y)}{m^2_{\pi_0} M_z(y)} \\
Y_{\pi Qu}& =& i \frac{(Y^{5D}_u)_{ij}}{\sqrt{2k}} \int_0^{y_1} dy
e^{-3\sigma(y)} Q^{m,i}_{L}(y)u^{0,j}_{R}(y)\frac{X_\pi'(y)}{m^2_{\pi_0} M_z(y)} \\
Y^\pi_1& =& i \frac{(Y^{5D}_u)_{ij}}{\sqrt{2k}} \int_0^{y_1} dy
e^{-3\sigma(y)}  Q^{m,i}_{L}(y)U^{n,j}_{R}(y) \frac{X_\pi'(y)}{m^2_{\pi_0} M_z(y)} \\
Y^\pi_2& =& i \frac{(Y^{5D^*}_u)_{ij}}{\sqrt{2k}} \int_0^{y_1} dy
e^{-3\sigma(y)} Q^{m,i}_{R}(y)U^{n,j}_{L}(y)\frac{X_\pi'(y)}{m^2_{\pi_0} M_z(y)} \, ,
\eea
and
\bea
a_{\pi qQ}&=& \frac{g^{5D}_{L}}{\sqrt{k}} \int_0^{y_1} dy e^{-2\sigma(y)}
q^{0,i}_{L}(y)Q^{n,j}_{R}(y)
\frac{X_\pi(y)}{m^2_{\pi_0}} \\
a_{\pi uU}&=&\frac{g^{5D}_{R}}{\sqrt{k}} \int_0^{y_1} dy e^{-2\sigma(y)}
u^{0,i}_{R}(y)U^{n,j}_{L}(y)  \frac{X_\pi(y)}{m^2_{\pi_0}}  \\
a_{\pi QQ}&=&\frac{g^{5D}_{L}}{\sqrt{k}} \int_0^{y_1} dy e^{-2\sigma(y)}
Q^{m,i}_{L}(y)Q^{n,j}_{R}(y)  \frac{X_\pi(y)}{m^2_{\pi_0}}  \\
a_{\pi UU}&=&\frac{g^{5D}_{R}}{\sqrt{k}} \int_0^{y_1} dy e^{-2\sigma(y)}
U^{m,i}_{R}(y)U^{n,j}_{L}(y)  \frac{X_\pi(y)}{m^2_{\pi_0}} 
\eea
where the $g^{5D}_{L,R}$ coupling are given by
\bea
g^{5D}_{L} =  {g^{5D}\over \cos \theta_W } \left(T_3 - Q_q \sin^2 \theta_W  \right)
\eea
\bea
g^{5D}_{R} = {g^{5D}\over \cos \theta_W} Q_q \sin^2 \theta_W ,
\eea
with $Q_q$  the charge of the quark, (here ${2 \over 3}$),
$\theta_W$ the weak angle and $T_3 = {1\over 2}$.
Note that when the interaction originates in the 5D Yukawa
couplings, the profile to use is the one coming from the CP-odd
Higgs component, i.e proportional to $X_\pi'(y)$. When the interaction
originates in the gauge fermion coupling and thus comes from the $Z_5$
component, the profile to use is proportional to $X_\pi(y)$, with
$X_\pi$ being the solution of Eq.~(\ref{Xeq}), using the decompositions
of Eqs.~(\ref{Z5decomp}) and (\ref{Pidecomp}). 

When the fermion matrix in (\ref{mumatrix}) is diagonalized, the coupling matrix
of fermions with the CP-odd field $\Pi_0(x)$ in (\ref{Ypimatrix}) is rotated,
and we can then extract all the physical Yukawa couplings. All these
couplings are needed later in order to compute the radiative
couplings of $\Pi_0(x)$ with gluons and photons.

\begin{figure}[t]
\center
\begin{center}
  \includegraphics[height=7cm]{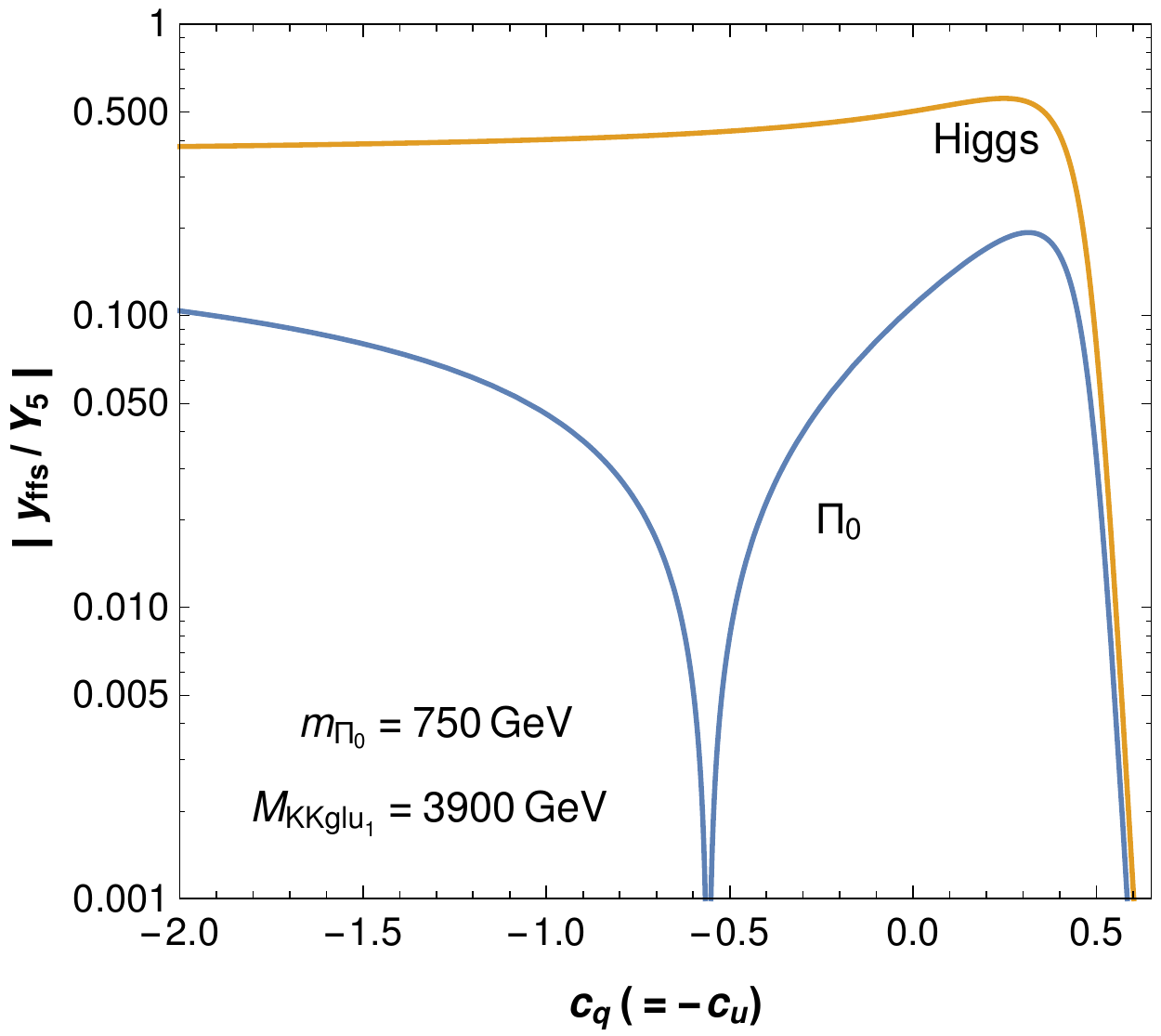}\ \ 
  \includegraphics[height=7cm]{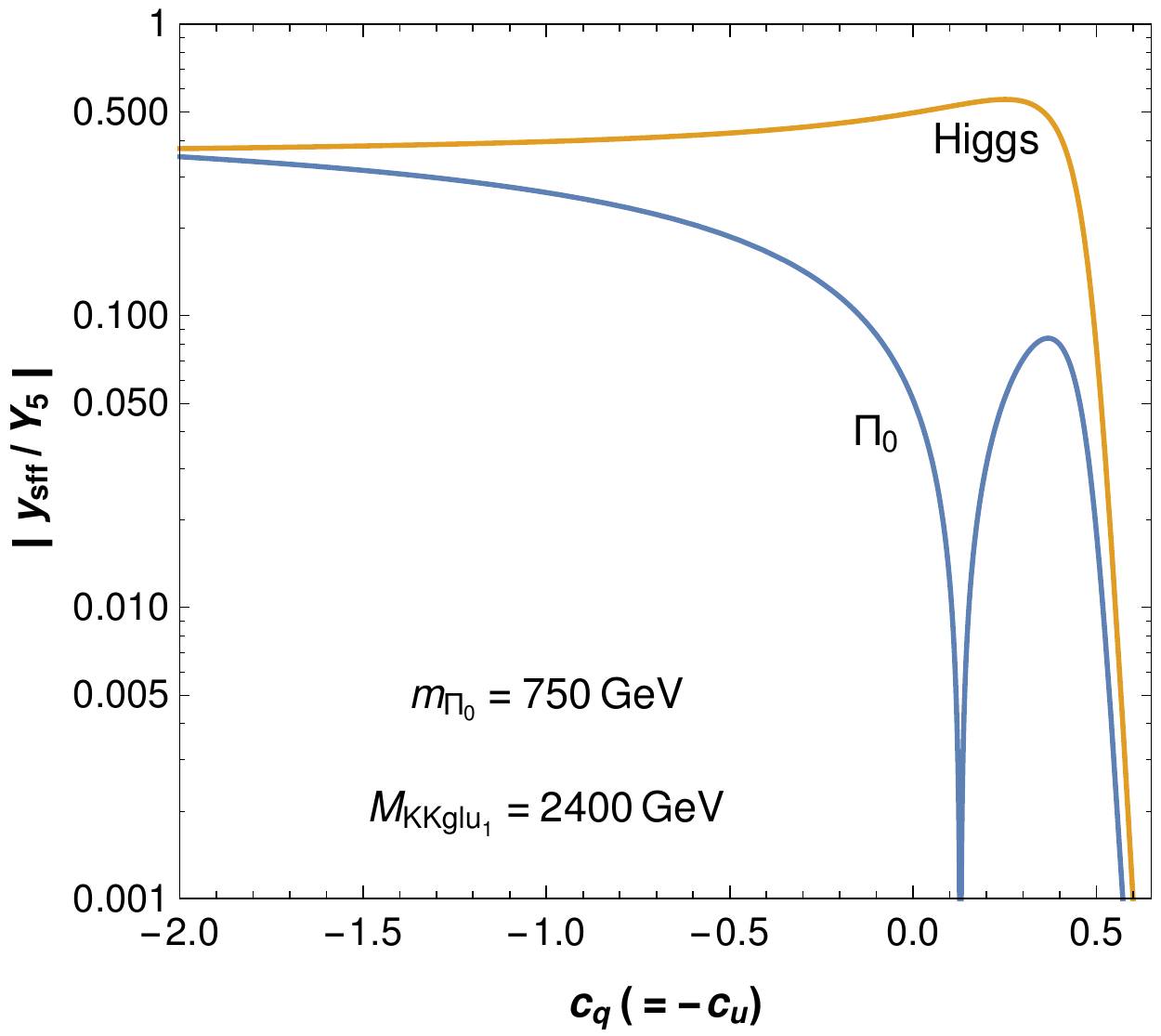}
  \includegraphics[height=7cm]{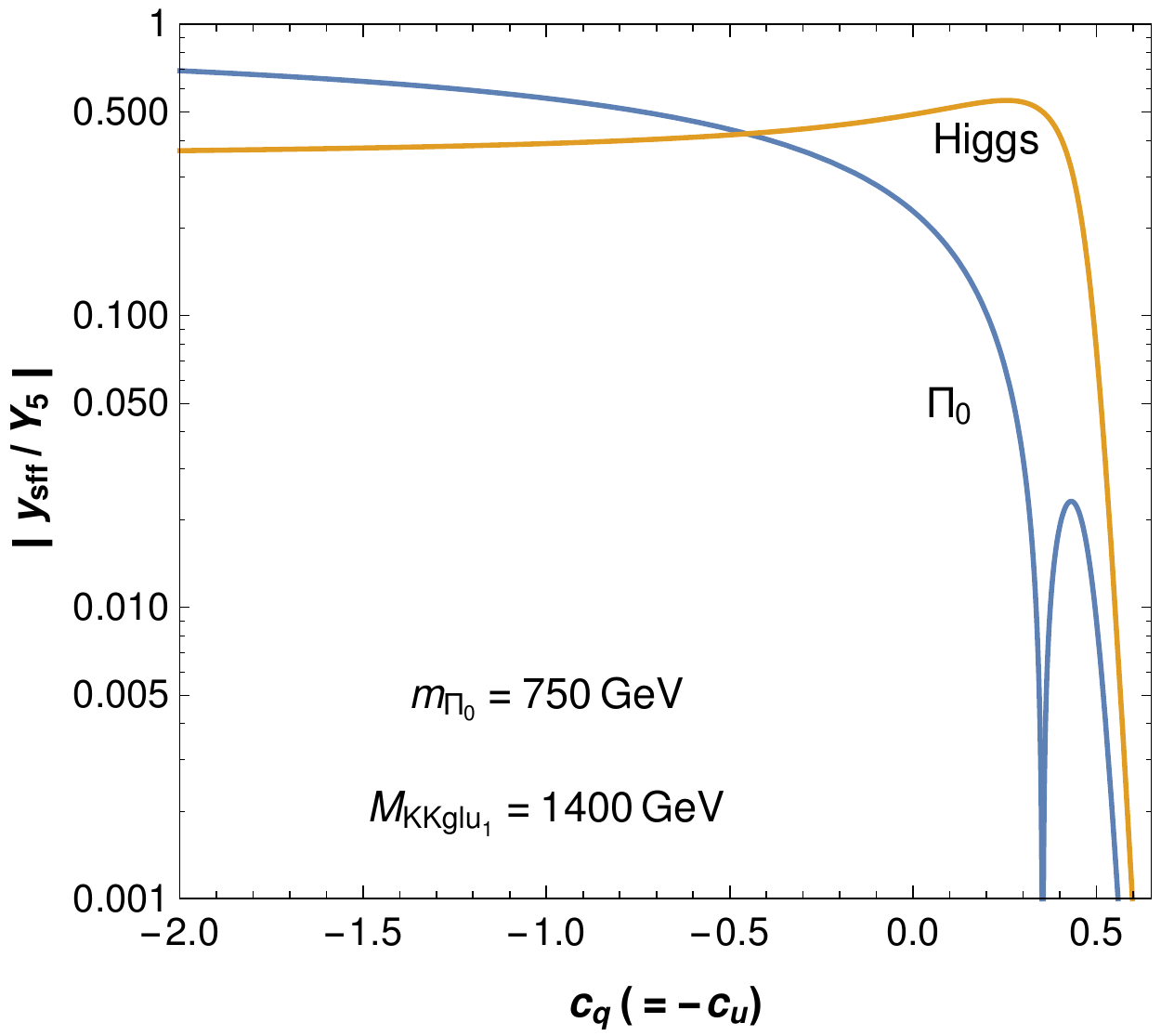}\ \ 
  \includegraphics[height=7cm]{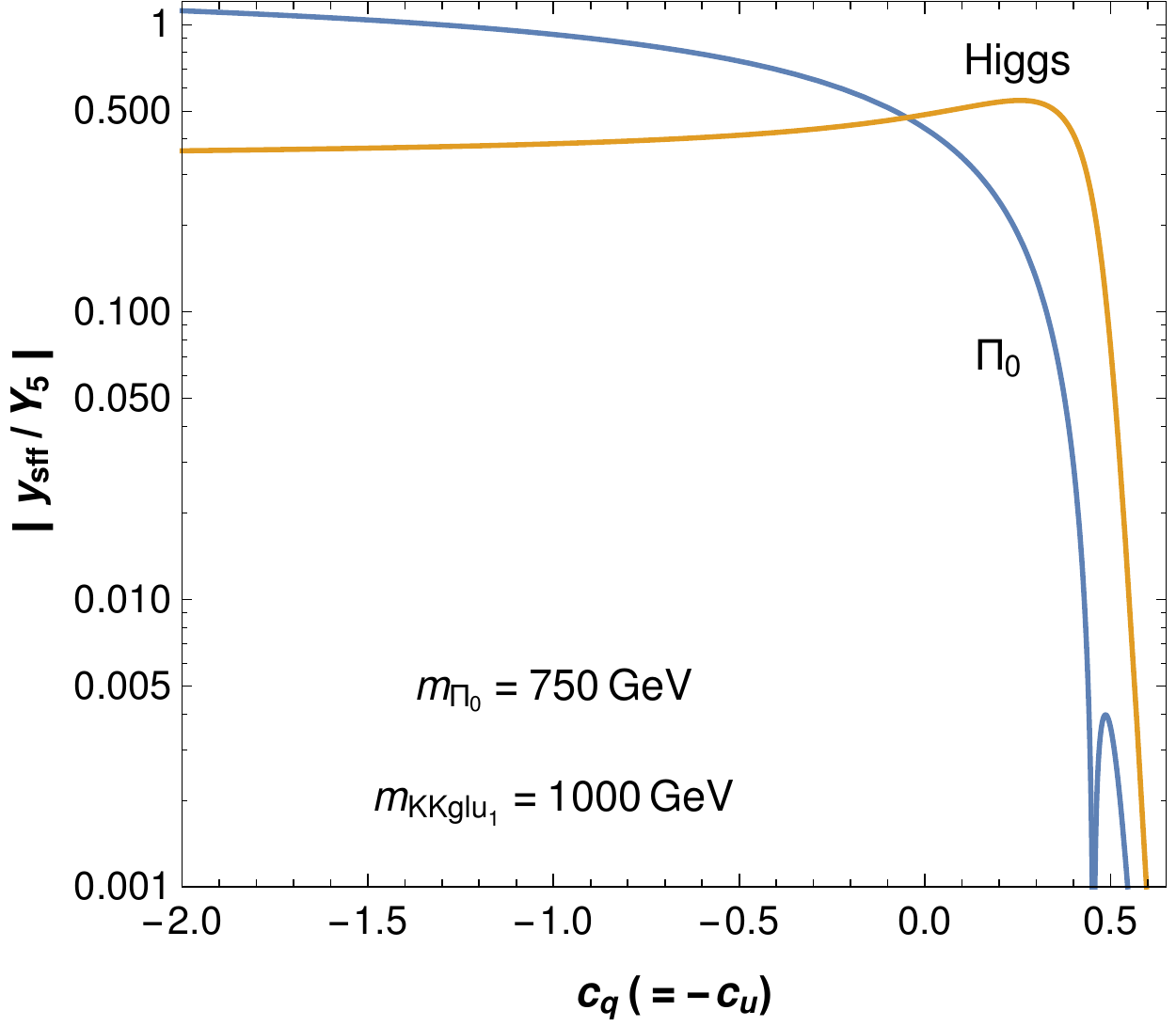}
\end{center}
\vspace{-.2cm}
\caption{Yukawa couplings between zero-mode fermions and the two
  lightest neutral scalars of the
  scenario, the $125$ GeV Higgs and the $750$ GeV CP-odd $\Pi_0$. The
  couplings are evaluated relative to the 5D bulk Higgs Yukawa coupling $Y_5$
  and are shown as a function of the fermion bulk mass parameter $c_q$ (in the case
  $c_q=-c_u$ for simplicity) for different overall KK scales, 
  $M_{KKglu_1}=3900$ GeV (upper left panel), $M_{KKglu_1}=2400$ GeV (upper right panel), $M_{KKglu_1}=1400$ GeV (lower left panel)  and $M_{KKglu_1}=1000$ GeV
  (lower right panel). The CP-odd scalar mass is set to $750$ GeV and
  for certain values of $c_q$ its Yukawa coupling to top quarks can be highly
  suppressed for typical top-quark values of the $c_i$'s.   } 
\label{fig:yPiplot}
\vspace{-.2cm}
\end{figure}

We first analyze the very important Yukawa
coupling between $\Pi_0(x)$ and top quarks, as this coupling might
dominate the decays of the CP-odd scalar.
This coupling comes essentially from the entry $(y^0_{\pi qu})_{ 33}$
(before rotation to the physical basis) although it receives small
corrections after going to the physical basis. We focus
 on $(y^0_{\pi qu})_{ 33}$ which comes from the overlap integral
\bea\label{YPtt}
(y^{0}_{\pi qu})_{33} &=&  i \frac{(Y^{5D}_u)_{33}}{\sqrt{2k}} \int_0^{y_1} dy
e^{-3\sigma(y)} q^{0}_{t}(y)u^{0}_{t}(y) \frac{X_\pi'(y)}{m^2_{\pi_0}
  M_z(y)}
\eea
In the RS limit, the warp factor is $\sigma(y)=ky$, and the top
profiles are $q^{0}_{t}(y)= f(c_q) e^{(2-c_q)ky}$ and
$u^{0}_{t}(y)=f(-c_u) e^{(2+c_u)ky} $, where $f(x)$ is a normalization
factor. We also have
$M_z(y)=\frac{g_5}{2c_W} v_0 e^{(a-1)ky}$, with $v_0$ a constant factor, so that the previous overlap
integral in this limit reads
\bea
(y^{0}_{\pi qu})_{33} =  i \frac{(Y^{5D}_u)_{33}}{\sqrt{2k}}
\frac{2 c_W}{g_5}\frac{f(c_q) f(-c_u)}{v_0 m^2_{\pi_0}}\int_0^{y_1} dy\
e^{(2-a-c_q+c_u)k y} X_\pi'(y)
\eea
We integrate this by parts to find
\bea
\!\!\!(y^{0}_{\pi qu})_{33} &=&  -i \frac{(Y^{5D}_u)_{33}}{\sqrt{2k}}
\frac{2 c_W}{g_5}\frac{f(c_q) f(-c_u)}{v_0
  m^2_{\pi_0}}\int_0^{y_1} dy\ (2-a-c_q+c_u) e^{(2-a-c_q+c_u)k
  y} X_\pi(y) + BT\ \ \ \ \ \
\eea
where
$BT= i \frac{(Y^{5D}_u)_{33}}{\sqrt{2k}} \frac{2 c_W}{g_5}\frac{f(c_q)
  f(-c_u)}{v_0 m^2_{\pi_0}}\  e^{(2-a-c_q+c_u)k y}
X_\pi(y)\Big|_0^{y_1} $ is a boundary term. Note that the
profile $X_\pi(y)$ has vanishing boundary conditions in the absence of
Higgs localized brane kinetic terms. In that limit we can
see that the coupling of the CP-odd scalar can actually vanish, when
$(2-a-c_q+c_u)=0$  \cite{Archer:2012qa}.
Note also that the Higgs localizer parameter $a$ is, in this RS limit, 
$a \gsim 2$ and the bulk parameters $c_q$ and $c_u$ are defined such
that, for example, charm or bottom quarks are assigned values more or
less $c_q \in (0.45,0.55)$ and $c_u\in 
(-0.5,-0.6) $, whereas for top we have $c_{q_3}\sim 0.45$ and $c_{u_3}> -0.45$.
This means that we expect the term $(2-a-c_q+c_u)$ to vanish, in the
limit of $a\sim 2$, when $c_q-c_u \sim 0$, so that the suppression in this
case seems only possible for the top quark, where both $c_q$ and $c_u$ could be small. 

Of course when the metric background is modified, and the boundary
conditions include the brane kinetic terms there will be deviations
from the previous values. Nevertheless it is clear that the Yukawa coupling
 of the CP-odd scalar field to top quarks can have highly
suppressed values. Another way to see this is to consider the overlap
integral in Eq.~(\ref{YPtt}). Because the profile $X_\pi(y)$ vanishes at the
boundaries (or $almost$ vanishes, for small brane kinetic terms),
then its derivative $X'_\pi(y)$ will have a node in the bulk, and
therefore will change sign. That means that there can be some
parameter choice for which it is possible for the overlap integral to
vanish, since the fermion zero mode profiles have no nodes 
in the bulk.

This feature is clearly seen in Fig. \ref{fig:yPiplot},
where we plot the absolute value of the Yukawa couplings between
zero-mode fermions and both the Higgs and the CP-odd scalar
$\Pi_0(x)$.\footnote{We are actually plotting the values defined
  in Eqs. (\ref{YYPI}) and (\ref{YYY1KK}), i.e. the zero mode Yukawa couplings before
  going to the fermion mass basis. In that basis, the
  couplings will inherit a small correction due to mixing with heavy
  KK fermions \cite{Azatov:2010pf}, so that the exact
  cancellation of the coupling will be replaced by a strong suppression.} The
couplings shown are relative to the 5D bulk Higgs Yukawa coupling
$Y_5$ and are plotted as  functions of the fermion bulk mass parameter
$c_q$  and $c_u$ (for the case where we take $c_q=-c_u$, for simplicity), for different overall KK scales. We observe that the
CP-odd Yukawa couplings are fairly similar to the Higgs Yukawa
couplings (i.e. exponentially sensitive to UV localization and then
top-like when the zero mode is IR localized) except that there is a
range of parameters where the coupling vanishes. Interestingly enough, 
this suppression happens for preferred values of the top quark bulk mass
parameters. This means that the existence of suppressed couplings to top quarks of the
CP-odd $\Pi_0$ is a natural possibility in this scenario.

\subsection{Radiative couplings to photons and gluons}
\label{subsec:gggamgam}

Just like in the Higgs boson case, the radiative couplings of $\Pi_0(x)$ to
gluons and photons will depend on the physical Yukawa couplings  
$y_{nn}$ between $\Pi_0$ and the fermions (zero modes and KK modes)
running in the loop, as well as on the fermion masses $m_n$ (the
eigenvalues of the mass matrix in Eq.(\ref{mumatrix})). The real
and imaginary parts of the couplings are associated with different loop functions, $A^S_{1/2}$ and
$A^P_{1/2}$, as they generate the two operators
$\Pi_0G_{\mu\nu}G^{\mu\nu}$ and $\Pi_0
G_{\mu\nu}\tilde{G}^{\mu\nu}$.\footnote{The Yukawa couplings of $\Pi_0$
  are mostly imaginary and thus the dominant contribution will come, as
  expected, from the operator $\Pi_0
  G_{\mu\nu}\tilde{G}^{\mu\nu}$. Still, small real Yukawa coupling
  components are generated when going to the fermion mass basis, and
  so  we keep the general formalism in our formulas.}

The production cross section through gluon fusion is %\cite{Gunion:1989we}
\bea
\sigma_{gg\rightarrow \Pi_0} = {\alpha_s^2 m_{\Pi_0}^2\over 576 \pi}
\left[\Big|\sum_{quarks}c^S_n\Big|^2 + \Big|\sum_{quarks}
  c^P_n\Big|^2\right]\  
\eea
and the decay widths to gluons and photons are
\bea
\Gamma_{\Pi_0\to gg} = \frac{\alpha_s^2
  m^3_{\Pi_0}}{54\pi^2}\frac{1}{v^2}\left[\Big|\sum_{{quarks
  }}c^S_n\Big|^2 + 
  \Big|\sum_{{quarks}} c^P_n\Big|^2\right]\  
\eea
\bea
\Gamma_{\Pi_0\to \gamma\gamma} = \frac{\alpha^2 m^3_{\Pi_0}}{192\pi^3}\frac{1}{v^2}
\left[\Big|\sum_{{quarks \atop leptons}} N_c Q_n^2 c^S_n\Big|^2 +
  \Big|\sum_{{quarks \atop leptons}} N_c Q_n^2 c^P_n\Big|^2\right]\  
\eea
where $\alpha_s$ and $\alpha$ are the strong and weak coupling
constants, $N_c$ is the number of colors and $Q_n$ is the charge of the fermion, and where
\bea
 c^S_n =\re\left({\frac{y_{nn}}{m_{n}}}\right) A^S_{1/2}(\tau_n)
\hspace{.6cm} {\rm and} \hspace{.6cm}
c^P_n = \im\left(\frac{y_{nn}}{m_{n}}\right) A^P_{1/2}(\tau_n)
\eea
with $\ \tau_n = m^2_{\Pi_0}/4m^2_n\ $ and with the loop functions
defined as \cite{Gunion:1989we}
\bea\label{eq:loopfunction}
A^S_{1/2}(\tau) &=& {3\over2}\left[\tau + (\tau -1)f(\tau)\right]\tau^{-2}, \\  
A^P_{1/2}(\tau) &=&- {3\over2} f(\tau)/\tau ,
\eea
and with
\bea
f(\tau) = \left \{
  \begin{tabular}{cc} $\left[\arcsin\sqrt{\tau}\right]^2$ &  $(\tau\le 1)$\\
  		 $ -{1\over4}\left[\ln\left({1+
        \sqrt{1-\tau^{-1}}\over 1-
        \sqrt{1-\tau^{-1}}}\right)-i\pi\right]^2$&  $(\tau> 1)$ .
\end{tabular}
\right.
\eea

For heavy KK quarks with masses $m_n$ much greater than the CP-odd mass
$m_{\Pi_0}$ (i.e. when $\tau$ is very small) the loop functions are essentially
constant, as they behave asymptotically as $\displaystyle \lim_{\tau
  \to 0} A^S_{1/2} =1 \hspace{.2cm} \mbox{and} \hspace{.2cm}
\lim_{\tau \to 0} A^P_{1/2} =3/2.$ On the other hand, for light quarks (all the SM quarks
except top and bottom), the loop functions essentially vanish 
asymptotically  as $\displaystyle \lim_{\tau \to \infty} A^S_{1/2}
= \lim_{\tau \to \infty} A^P_{1/2} = 0.$

\begin{figure}[t]
\center
\begin{center}
  \includegraphics[height=7cm]{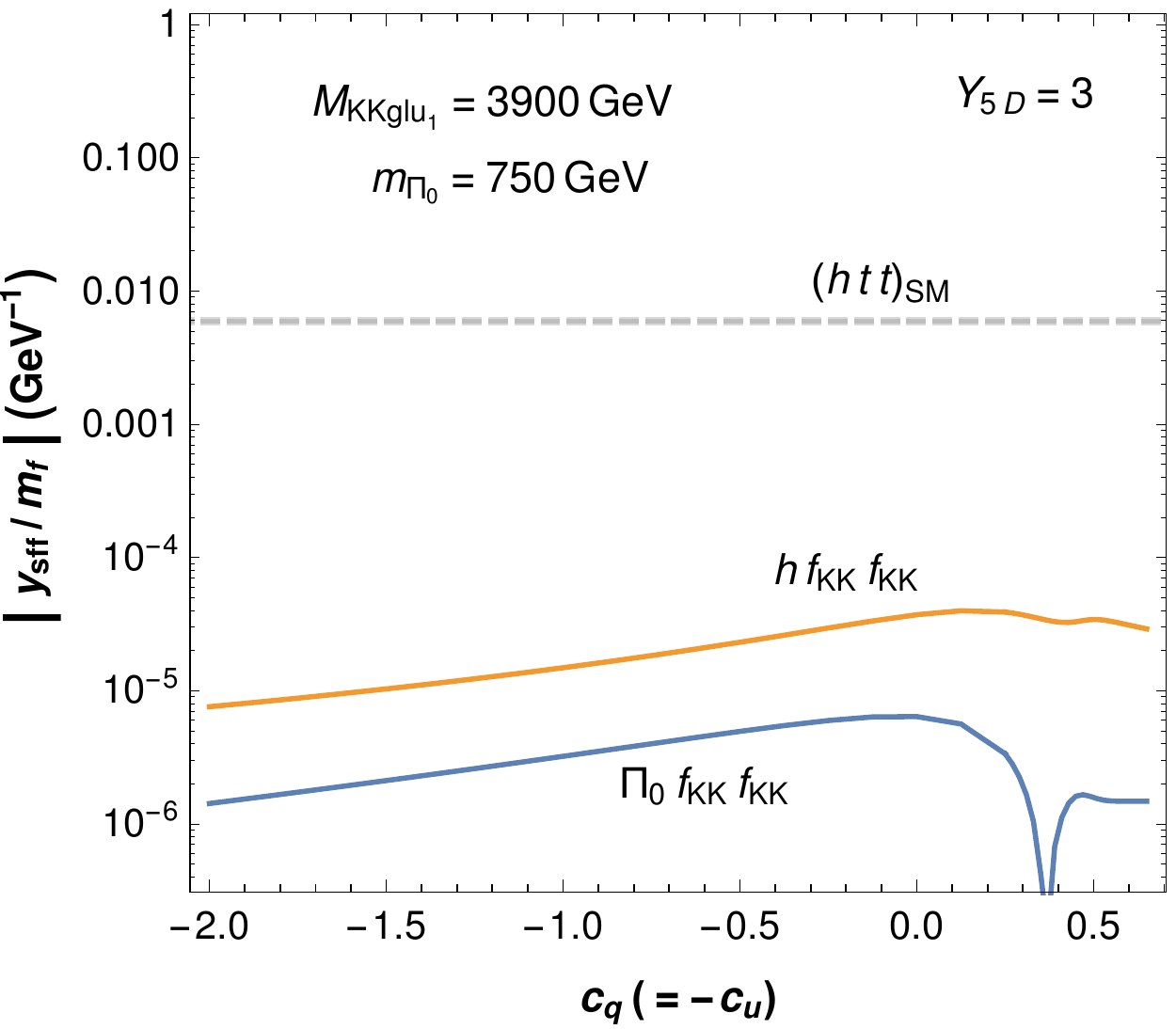}\ \ 
  \includegraphics[height=7cm]{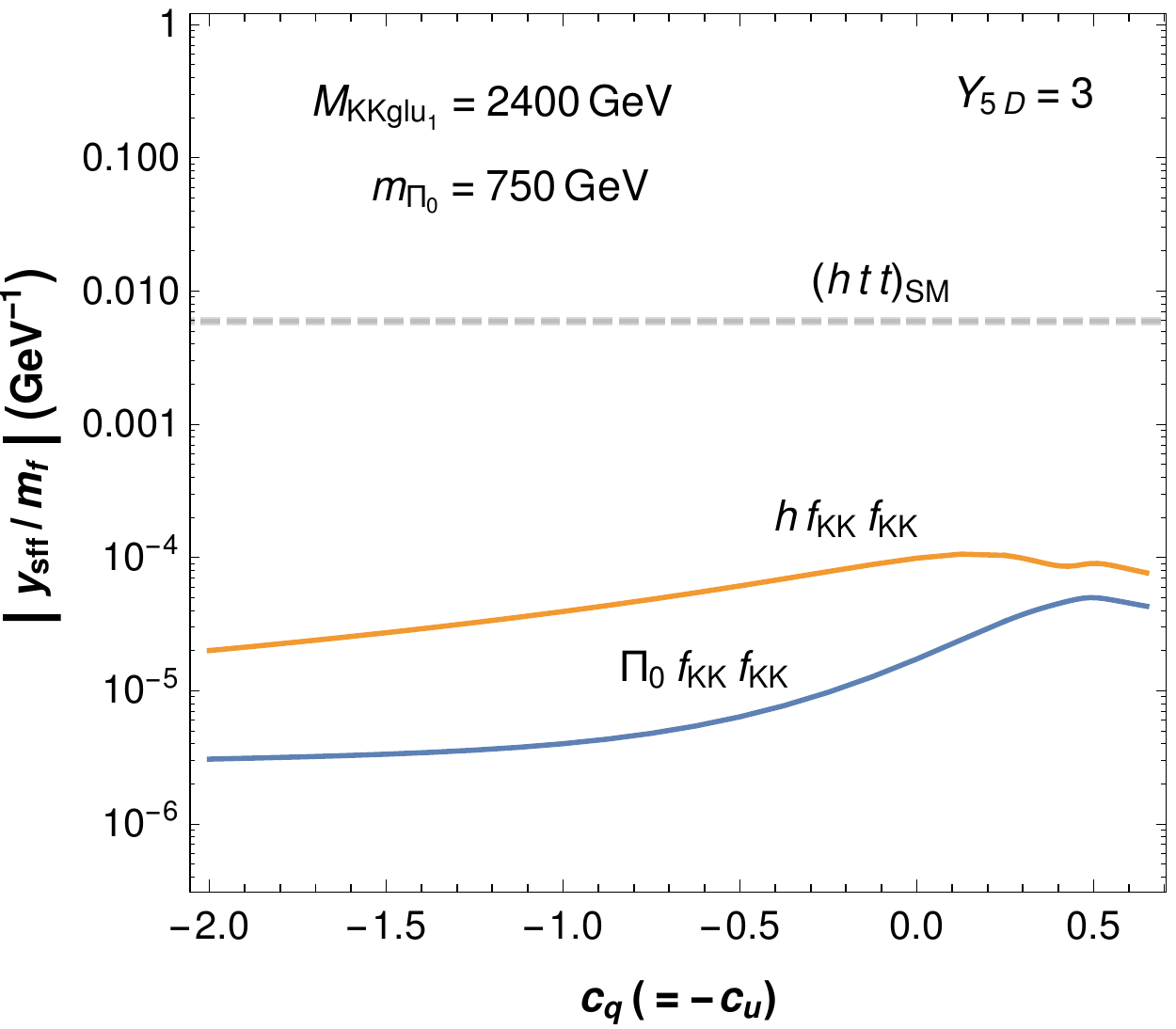}
  \includegraphics[height=7cm]{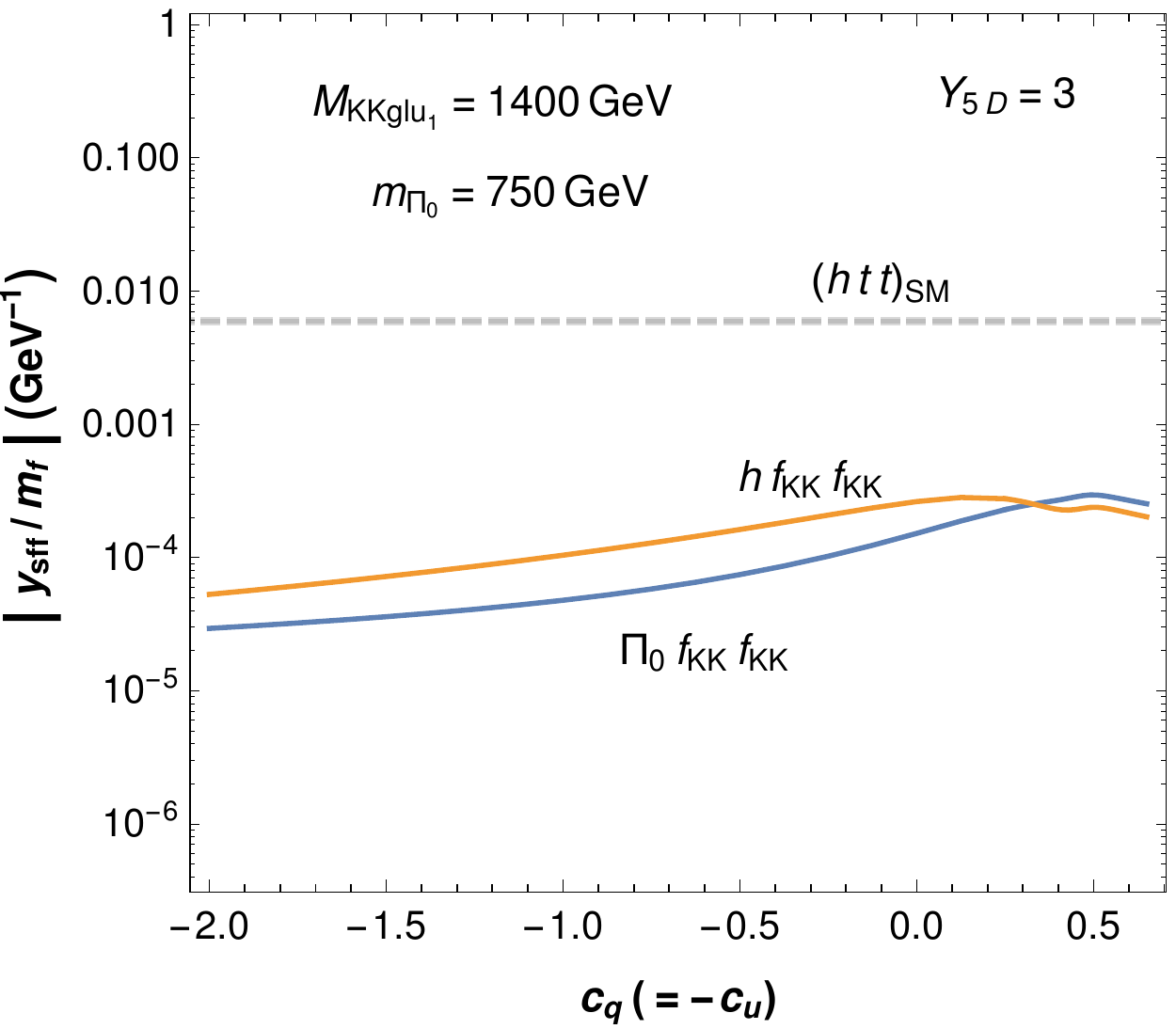}\ \ 
  \includegraphics[height=7cm]{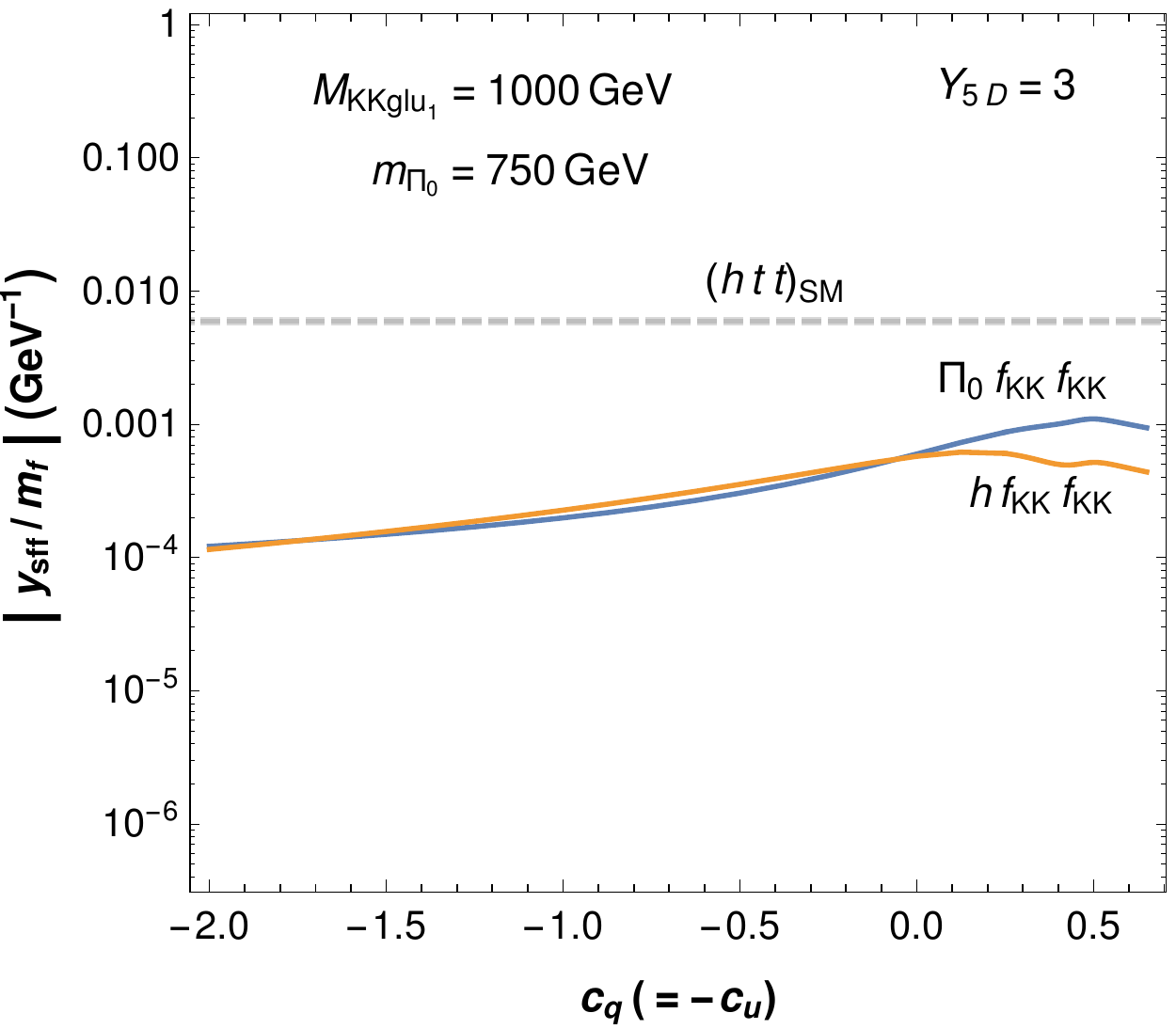}
\end{center}
\vspace{-.2cm}
\caption{Yukawa couplings between the lightest KK fermion and the $125$ GeV
  Higgs (middle curves) and the $750$ GeV CP-odd $\Pi_0$ (lower curves),
  divided by the KK fermion mass, for different values of the lightest KK gluon mass $M_{KKglu_1}$, as indicated on the panels. This mass normalized Yukawa coupling
  gives an estimate of the relative contribution of the respective KK
  fermion to the radiative coupling of the scalar to gluons and
  photons, to be compared with the mass normalized SM
  coupling of Higgs to top quarks (shown as a dashed line).}
\label{fig:yPiKKplot}
\vspace{-.3  cm}
\end{figure}

Moreover, we investigate a parameter region where the couplings
of $\Pi_0$ to top quarks are highly suppressed. This means that the
production mechanism must rely exclusively on the heavy KK fermions running in the
loop and as we have seen, this coupling depends on the ratio $\frac{y_{nn}}{m_n}$ between
the physical Yukawa coupling and the mass of the fermion running in
the loop. To have an idea of the relative contribution of
each of these KK fermions in the loop, in Fig. 
\ref{fig:yPiKKplot} we plot the mass normalized Yukawa couplings of Standard
Model Higgs with top quarks, Higgs with first KK fermion and of $\Pi_0$ to
first KK fermion, for different values of the KK scale.
As expected we see that the $c_q$ dependence is mild (i.e. all KK fermions
of any flavor will couple with similar strength) and also, as expected,
we observe that the mass normalized couplings are quite suppressed
with respect to the SM top quark case. Still the multiplicity of KK
fermions is high, since there are 6 families of quarks and 3 families of charged
leptons (the latter run in the diphoton loop), and for each family
there are a few KK levels that give important contributions to the rate.  

A numerical scan of the couplings, including all families and 3 full
KK levels is computationally too intensive, so in order to produce
the couplings plotted in Fig. \ref{fig:yPiKKplot} we  performed
an approximation,  sufficient for the purposes of the graph.

The KK fermion Yukawa couplings plotted neglect mixings between
different KK levels and different flavors, and with the zero mode fermions. They are obtained as follows.
 Consider the $2\times2$ KK mass matrix
\bea
\left(\begin{matrix}
 Q_L(x)\ U_L(x)
\end{matrix}\right)\  M_u\ \left(\begin{matrix}
  Q_R(x)\\ U_R(x)
\end{matrix}\right) , 
\eea
where $Q_L(x)$ and $Q_R(x)$ represent here a single flavor and a
single KK level of the vector-like KK up-type doublets, 
and similarly for $U_L(x)$ and $U_R(x)$, vector-like KK up-type singlets. 
The mass matrix is thus
\bea
M_u= \left(\begin{matrix}
m_Q & Y_1 \\
Y_2 &  m_U
\end{matrix}\right),
  \label{QUmatrix}
\eea
where the diagonal entries are the KK masses (large) whereas the off-diagonal
entries are coming from Yukawa couplings and are therefore smaller.
In order to give a simple estimate, we take for simplicity the fermion
bulk mass parameters as $c_q=-c_u$, and the bulk Higgs Yukawa $Y^{5D}$
to be real, which leads to $Y_1=Y_2$ and $m_Q=m_U=m_{KK}$, with the masses
and profiles obtained by solving Eq. (\ref{fermionprofile}). With
the KK fermion profiles one obtains the off-diagonal entries 
\bea
Y_1= \frac{(Y^{5D}_u)}{\sqrt{k}} \int_0^{y_1} dy e^{-4\sigma(y)}
\frac{ v(y)}{\sqrt{2}} Q_{L}(y)U_{R}(y) .
\eea
The matrix that diagonalizes (\ref{QUmatrix}) in this simple limit
($c_q=-c_u$) is $\left(\begin{matrix}
-\frac{1}{\sqrt{2}} & \frac{1}{\sqrt{2}}  \\
\frac{1}{\sqrt{2} }& \frac{1}{\sqrt{2}}
\end{matrix}\right)$ and the eigenvalues are $m_1=m_{KK}-Y_1$ and $m_2=m_{KK}+Y_1$.

Now we apply this rotation to the CP-odd Yukawa coupling matrix
\bea
Y_\Pi= \left(\begin{matrix}
{\cal O}(g) & Y_\pi \\
Y_\pi &  {\cal O}(g)
\end{matrix}\right),
\eea
where
\bea
Y_\pi& =& i \frac{(Y^{5D}_u)_{ij}}{\sqrt{2k}} \int_0^{y_1} dy
e^{-3\sigma(y)}  Q_{L}(y)U_{R}(y) \frac{X_\pi'(y)}{m^2_{\pi_0} M_z(y)},
\eea
and where for simplicity we have neglected gauge couplings terms compared to IR Yukawa
terms (safe assumption when $Y^{5D}$ is large).

After diagonalization, we obtain the two physical couplings between
$\Pi_0$ and the KK fermions. When we normalize the couplings by the
two eigenmasses and add the two contributions,\footnote{One needs to
  add the two contributions since there is a cancellation happening
  level by level.} we obtain
\bea
\sum_{i=1}^2 \frac{y_i}{m_i} =\frac{y_1^{\pi}}{m_1}+\frac{y_2^{\pi}}{m_2} = -2 \frac{Y_1
  Y_\pi}{m_{KK}^2-Y_1^2} \simeq -2\frac{Y_1  Y_\pi}{m_{KK}^2}.
\eea
The last expression corresponds to the {\it mass normalized} Yukawa couplings
of $\Pi_0$ plotted in Fig. \ref{fig:yPiKKplot}, and this describes
very closely the behavior of the couplings obtained in the full flavor calculation. 
The parametric dependence of these couplings is $Y_{5D}^2 v/m_{KK}^2$,
so that if $Y^{5D}\sim 3$ we expect mass 
normalized couplings of order $(10^{-3}-10^{-4})$ GeV$^{-1}$, if the
overlap integral is of ${\cal O}(1)$. Since all the profiles of the
integral are IR localized, one expects that integral to be ${\cal
  O}(1)$, although the precise numerical result varies  between
$0.5$ and $0.05$, depending on the values of the $c_q$ parameter, as
shown in the plots.

All in all it seems likely that after taking into consideration all the
fermion flavors, and for a KK scale of order $1-2$ TeV, the overall KK
fermion contribution to the radiative couplings of $\Pi_0(x)$ to photons
and gluons can be close to the top quark contribution to the
gluon and photon couplings of the Higgs in the SM model.  

\subsection{$\Pi_0Zh$ coupling}
\label{subsec:zh}

The coupling between the CP-odd scalar, the $Z$ boson and the Higgs will
be extracted from the kinetic operator of the 5D Higgs,
\bea
\int d^4x dy\ e^{-2\sigma} D_\mu H^\dagger D^\mu H \left(1+ \delta(y-y_i) \frac{d_i}{k}\right).
\eea
Expanding the SM-like Higgs mode using Eq. (\ref{Hexpansion}) as well as the
SM-like $Z_\mu$ and the $750$ GeV $\Pi_0$ using  Eqs. (\ref{Aexpansion}) and  (\ref{Piexpansion}), we can
obtain the coefficient $g_{\Pi hZ}$ of the operator
$Z^\mu(x)\Big(h(x)\partial_\mu\Pi_0(x)+\Pi_0(x)\partial_\mu h(x)\Big)$ 
\bea
g_{\Pi hZ} = \frac{g_5^2}{4c_W^2} \int dy e^{-2\sigma} v(y)h(y) f_z(y) \frac{X'(y)}{M^2_z(y) m^2_{\pi}} \left(1+ \delta(y-y_i) \frac{d_i}{k}\right)
\eea
Now since $M_z(y)=\frac{g_5}{2c_W}e^{-\sigma} v(y)$, 
$h(y)\sim v(y)/v_4$, and  $f_z\simeq 1/\sqrt{y_1}$ we can write
\bea
g_{\Pi hZ} \simeq \frac{1}{\sqrt{y_1}v_4 m^2_\pi} \big(X(y_1)+
X'(y_1)\frac{d_1}{k}\big)=0,
\eea
where we have used the boundary conditions for the profile $X(y)$ (see
Eq.(\ref{Xbc})) and assumed no UV brane kinetic term ($d_0=0$).

The coupling should thus vanish in the limit of flat $Z$ boson profile $f_z(y)$,
and when the nontrivial Higgs VEV $v(y)$ is proportional to the Higgs
scalar profile $h(y)$. Corrections to these limits scale as
$v_4^2/m_{KK}^2$ and $m_h^2/m_{KK}^2$ in the RS case, and so we expect the
overall coupling to be highly suppressed.

The partial width for the decay $\Pi_0 \to h Z$ is \cite{djouadi1996}
\bea
\Gamma(\Pi_0 \to h Z)= \frac{g^2_{\Pi hZ}}{16\pi}
\frac{m_Z^2}{m_{\pi}}
\sqrt{\lambda(m^2_h,m^2_Z;m^2_\pi)}\lambda(m^2_h,m^2_\pi;m^2_Z) 
\eea 
where $m_Z$, $m_\pi$ and $m_h$ are the masses of the particles involved and
where $\lambda(x,y;z) = (1 - x/z - y/z)^2 - 4 x y/z^2$.
With the masses  $m_Z=91$ GeV, $m_\pi=750$ GeV and $m_h=125$ GeV, the
width becomes $\Gamma(\Pi_0 \to h Z) \sim (900\ g^2_{\Pi hZ})$ GeV.

For example we compute numerically $g_{\Pi hZ}$ for three different values of
$M_{KKglu_1}$, and with $m_{\Pi_0}=750$ GeV and find
\begin{center}
\begin{tabular}{|c||c|c|c|c|}
\hline
$M_{KKglu_1}$ &\ 1000 GeV\ &\ 1400 GeV\ &\ 2400 GeV\   \\
\hline
$g_{\Pi hZ}$ & $3.8\times 10^{-4}$ & $3.1\times 10^{-3}$ &$1.1\times 10^{-2}$  \\
\hline
$\Gamma(\Pi_0 \to hZ$) &$1.3\times 10^{-4}$ GeV  &$8.8\times 10^{-3}$
GeV   &$0.11$ GeV \\
\hline
\end{tabular}
\end{center}
Note that the couplings and widths are small but 
 we observe that the partial width becomes larger as the KK
mass scale is increased. Since we need  the partial width
$\Gamma(\Pi_0 \to \gamma \gamma)$  to be similar to that of a $750$ GeV Higgs,
i.e. $\Gamma(h_{750GeV} \to \gamma \gamma) \sim 10^{-5}$ GeV, we expect that
at $M_{KKglu_1} =1400$ GeV the $Zh$ signal should start putting too
much pressure on the allowed parameter space. This is
confirmed in the full numerical analysis presented in the next section.

\subsection{Estimates and numerical results}
\label{subsec:numerical}
With all the previous ingredients one can  estimate the
viability of this scenario in terms of the possible diphoton excess.
Let's choose the KK scale such that $M_{KKglu_1}=1400$ GeV; when the
bulk mass parameters of the top quark are around $|c_{u_3}|\sim 0.35$ 
we know that the top quark will have highly suppressed couplings to
$\Pi_0$, as shown in the third panel of Fig. \ref{fig:yPiplot} . At the same
time, the couplings of the KK tops (as well as all other KK quarks)
will have relatively strong Yukawa couplings to $\Pi_0$ (third panel
of Fig. \ref{fig:yPiKKplot}), so that the contribution of each of them to the
radiative coupling of $\Pi_0$ to gluons is about an order of magnitude
smaller than the top contribution to the $h-glu-glu$ coupling of the
SM. Thus we could estimate that the overall contribution of all flavors
and KK excitations can make up for the suppressed coupling, so that
the production cross section of $\Pi_0$ is similar to that of a
$750$ GeV SM-like Higgs.

The production cross section of a $750$ GeV Higgs through gluon fusion,
at the LHC running at 13 TeV is $497$ fb 
  \cite{Dittmaier:2011ti}, so, roughly, this could be assumed for the $\Pi_0$ production cross section.

Since the $\Pi_0$ decays to top quarks and $hZ$ are suppressed in this parameters
space point, and its decays to $WW$ and $ZZ$ can only be radiative via
the CP-odd gauge boson kinetic operator,  the main
decay channel is into gluons so that the branching of the diphoton
channel should be very roughly
\bea
Br(\Pi_0 \to \gamma\gamma) \sim \frac{\alpha^2_{em}}{8 \alpha^2_s}
\frac{N_\gamma}{N_{glu}}
\eea
where $N_\gamma$ and $N_{glu}$ are the multiplicities of states running
in the $(\Pi_0 \gamma \gamma)$ loop and in the $(\Pi_0\ glu\ glu)$ loop respectively.

In the diphoton loop there are $3$ extra families of charged lepton KK
excitations making the multiplicity of states greater. If their 
multiplicity and their Yukawa couplings can partially make up for the color factor of 8,
then the diphoton cross section might become of ${\cal O}$ (1 fb), as hinted
by the December 2015 LHC data. 
 
To complete the analysis, we perform a full numerical computation of production and
branching ratios in a setup where we consider an effective 4D scenario
including three full KK levels for all fields, i.e., we consider $21 \times 21$ fermion
mass matrices, which we diagonalize in order to obtain the
physical Yukawa couplings. We choose a set of $c$-parameters and 5D
Yukawa entries such that the SM masses and mixings are reproduced; the
specific flavor choice for these parameters should not affect much the 
overall results since these depend on overlap integrals between IR
localized fields, with very loose $c$-dependence. We choose the
background metric parameters so that precision electroweak 
bounds are kept at bay, i.e. $\nu=0.5$ and $y_s=1.04 y_1$. Two average
5D Yukawa scales are considered, 
$Y_{5D} \simeq 3$ and $Y_{5D}\simeq 2$,  to show the dependence on this
parameter, and we also consider two different KK mass scales,
$M_{KKglu_1}=1000$ GeV and $M_{KKglu_1}=1300$ GeV, which turn out to
lead to successful signal generation. 

In order to see how tuned is the choice of top $c$-parameter, we plot the
production cross section of the CP odd resonance, followed by decays
into $\gamma \gamma$, $t\bar{t}$ and $Zh$,  
 as functions of $c_{u_3}$ (the bulk mass parameter of 
the 5D singlet top quark), with the doublet bulk mass 
parameter fixed at $c_{q_3}=0.4$. (This value ensures typically
  suppressed bounds from $Zb_L{\bar b}_L$ bounds \cite{Cabrer}).
\begin{figure}[t]
\center
\begin{center}
  \includegraphics[height=7cm]{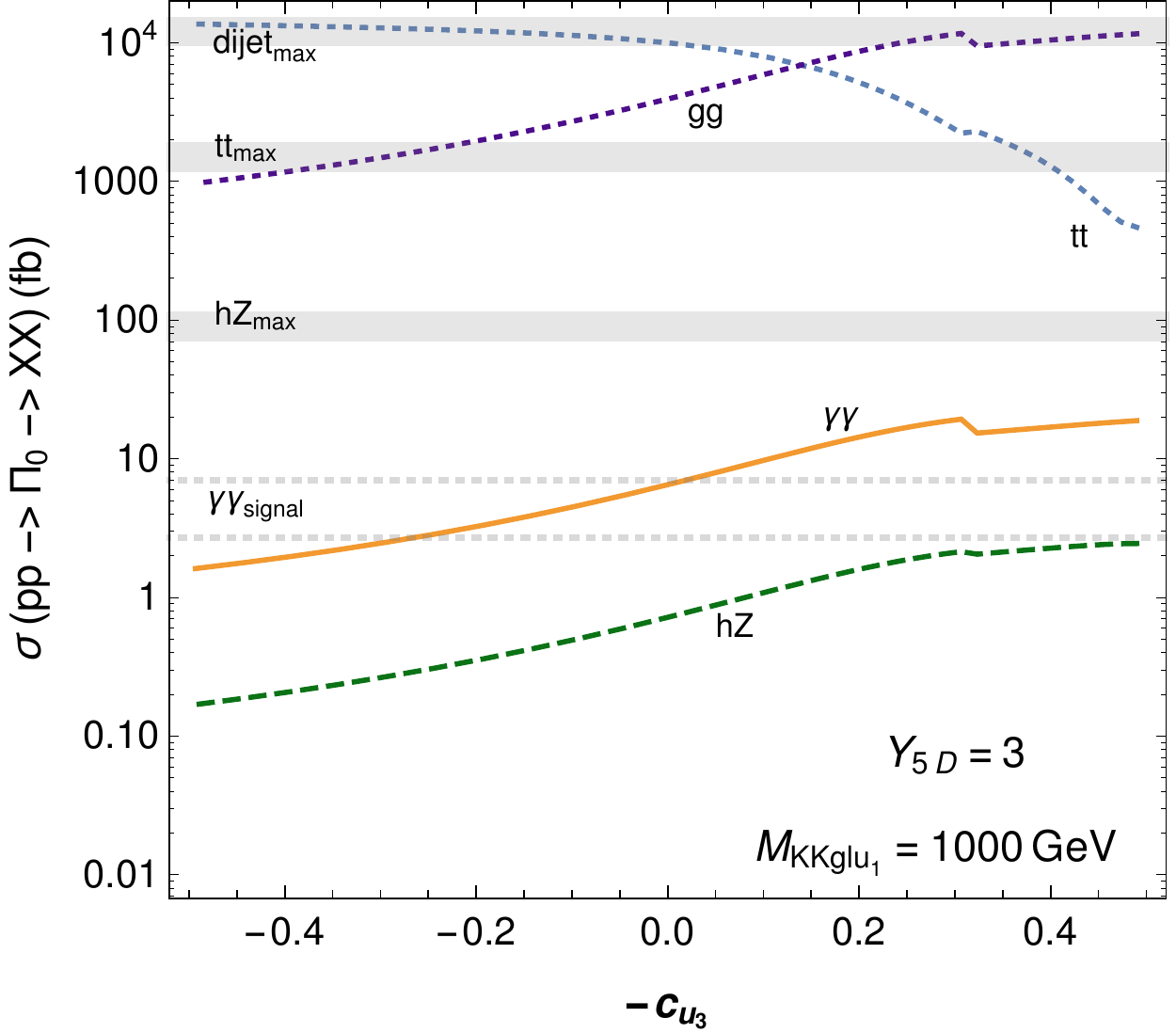}\ \ 
  \includegraphics[height=7cm]{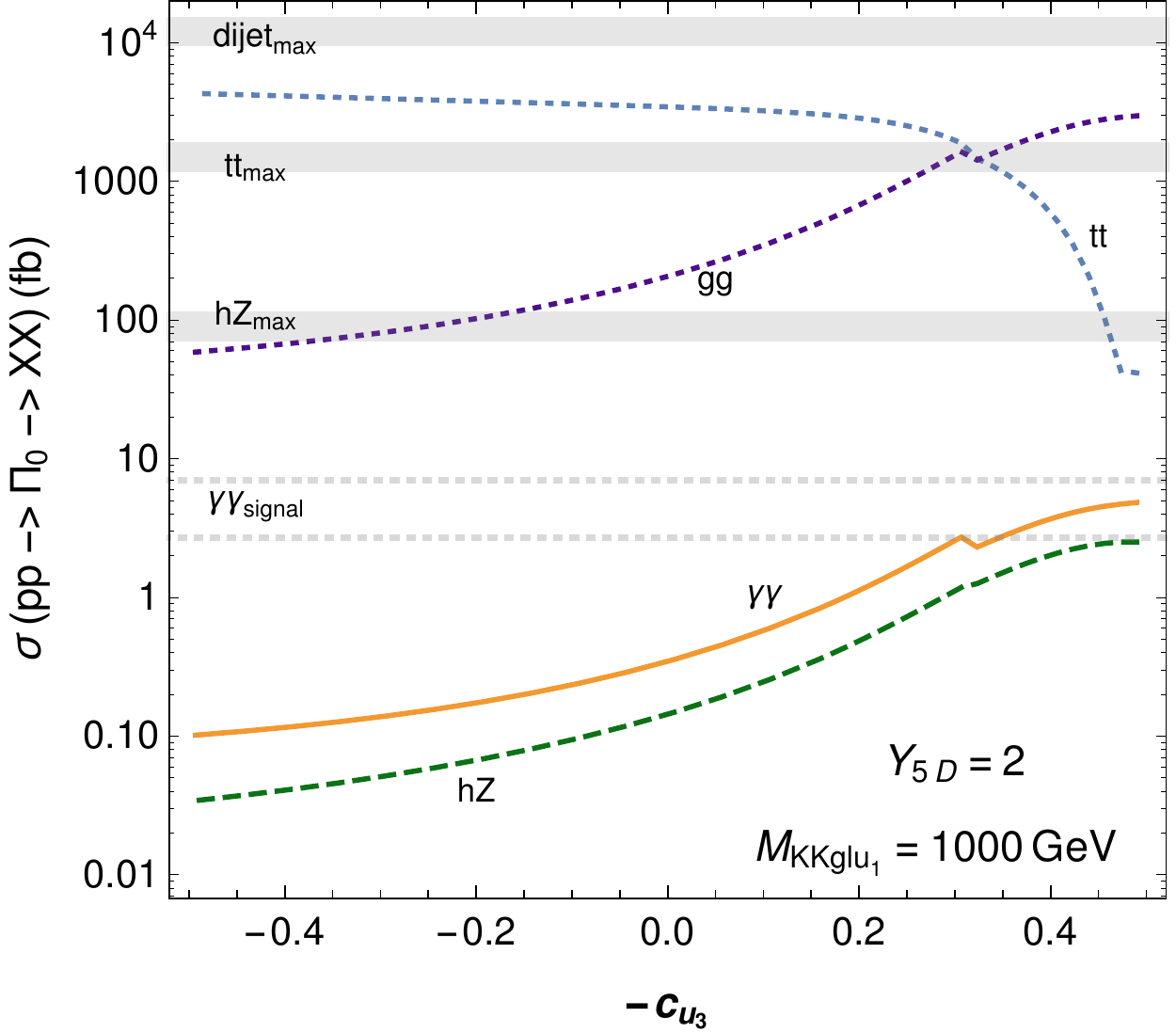}
  \end{center}
\vspace{-.2cm}
\caption{Production cross sections at the LHC running at 13 TeV in the
  channels $\gamma\gamma$, $t\bar{t}$ and $hZ$. The lightest KK boson
  mass is $1000$ GeV and the 5D Yukawa couplings are $Y_{5D}\simeq 3$  in  the left
  panel and  $Y_{5D}\simeq 2$  in the right panel. The signal can always be
  produced at this KK scale but for larger 5D Yukawa couplings there are
  either too many top pairs or too many dijets produced. Overall we
  can conclude that for $M_{KKglu_1} = 1000$ GeV and when $1\lsim Y_{5D}\lsim 3$ and $0.2\lsim |c_{u_3}|
  \lsim 0.5$ the signal is easily reproduced.}  
\label{fig:prodplot1}
\vspace{0.  cm}
\end{figure}\begin{figure}[t]
\center

\begin{center}
  \includegraphics[height=7cm]{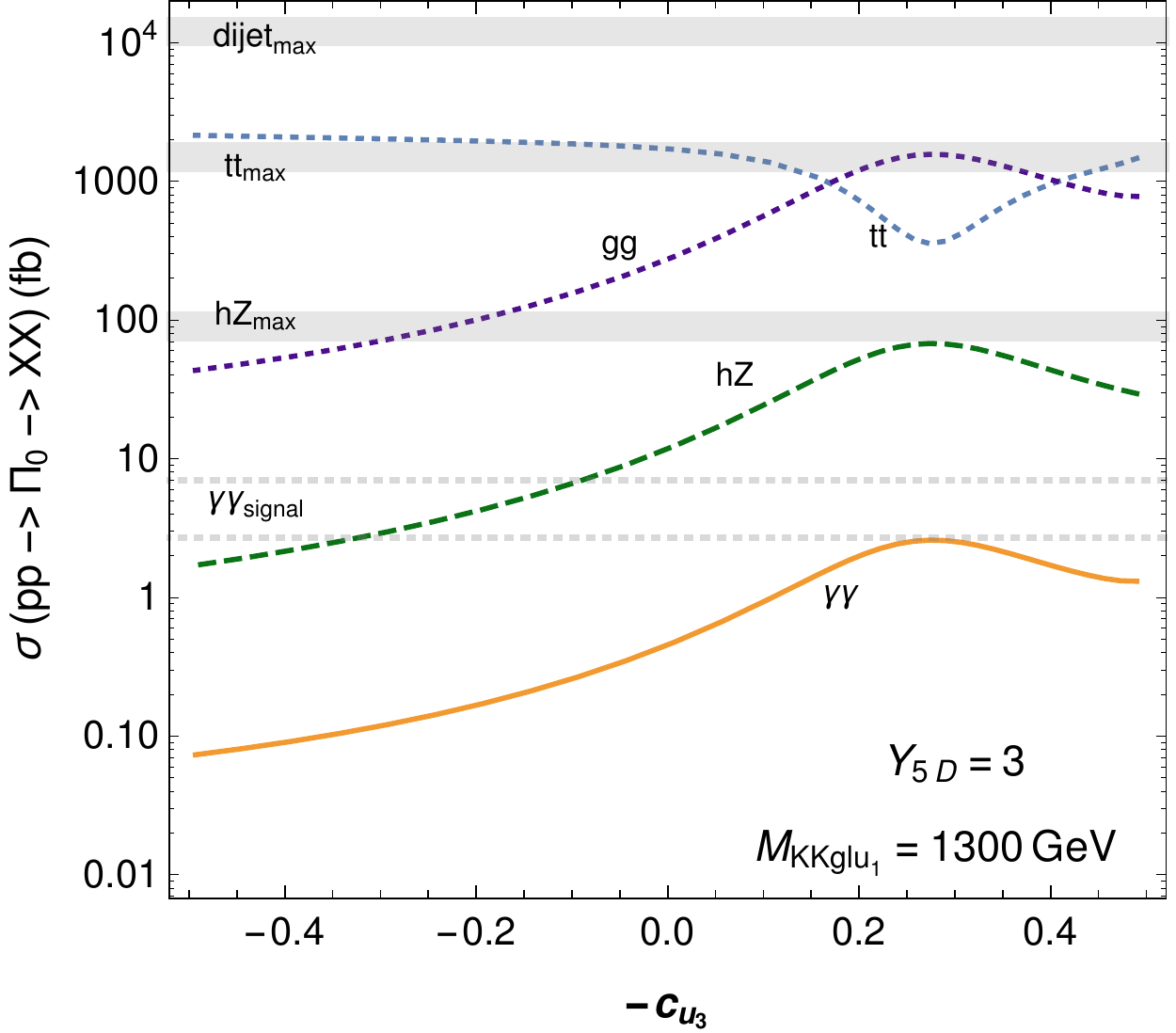}\ \ 
  \includegraphics[height=7cm]{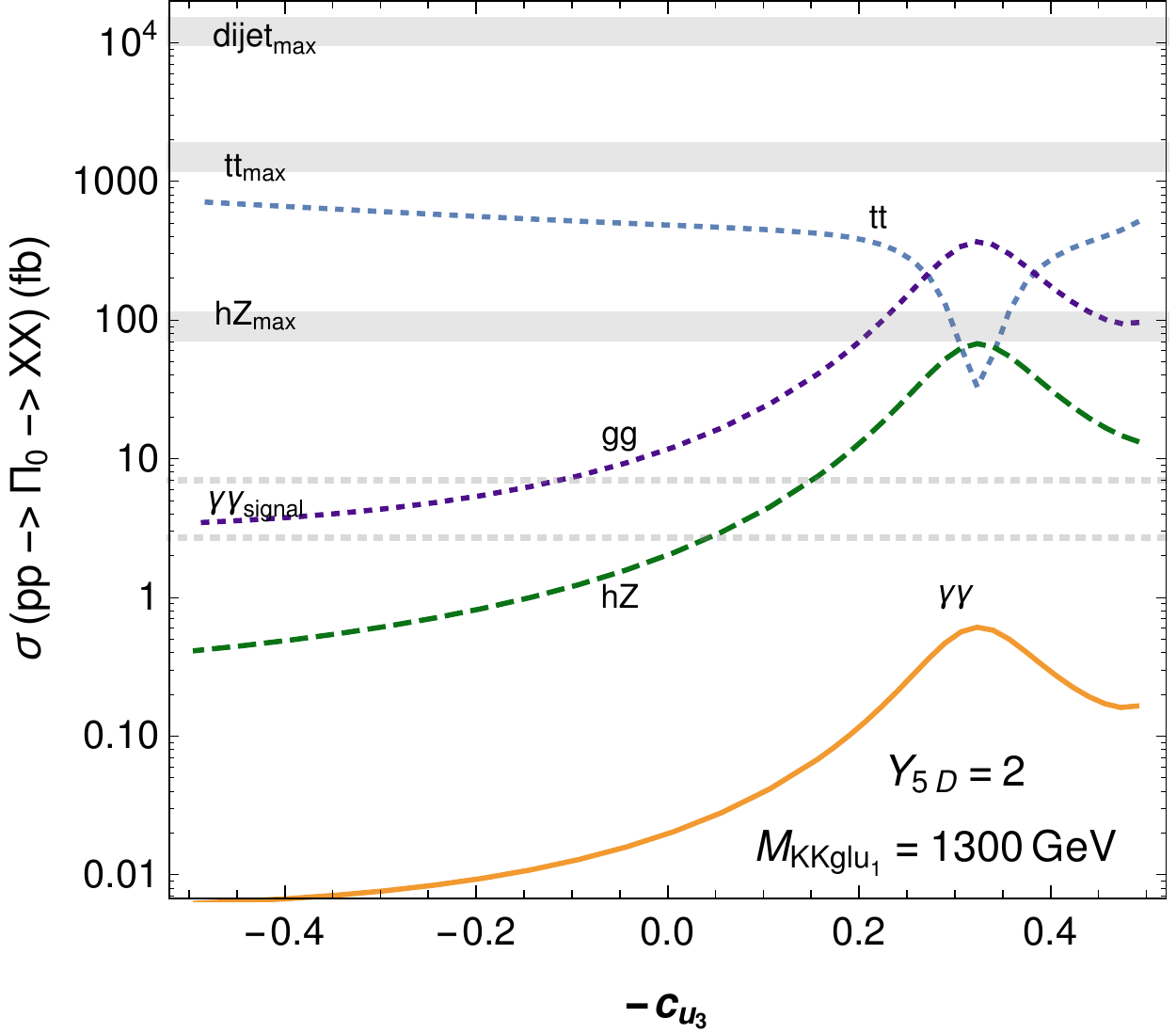}
  \end{center}
\vspace{-.2cm}
\caption{Same as the previous figure but for a KK scale of $1300$ GeV.
In this case the signal can only be achieved for larger Yukawa couplings, and
at a more localized region near $c_{u_3}\sim 0.25$. The bound from $hZ$
now becomes the main constrain. For heavier KK mass scales, the signal
can only be produced for much larger Yukawa couplings and the $hZ$ bound becomes
too constraining.}
\label{fig:prodplot2}
\vspace{0.  cm}
\end{figure}

The results are shown in Figs. \ref{fig:prodplot1} and
\ref{fig:prodplot2}, where we have chosen two KK scales, 
$M_{KKglu_1}=1000$ GeV and $M_{KKglu_1}=1300$ GeV. In both cases we show
results for $Y_{5D}\simeq 3$ and $Y_{5D}\simeq 2$ to illustrate the sensitivity
on this bulk parameter, crucial for enhancing the radiative couplings
of $\Pi_0$. When the KK scale is smaller and 5D Yukawa couplings are larger,
the production of top pairs and gluon pairs can be too large. By
reducing the 5D Yukawa couplings, enough signal can be generated with
the dijets and top pairs under control, as well as the $Zh$ decay. The
values of $|c_{u_3}|$ must be in the region  $0.2\lsim |c_{u_3}|
\lsim 0.5$. For slightly smaller KK scales one expects a similar
behavior, but such that 5D Yukawa couplings should be smaller in order to
suppress overproduction of $\Pi_0$ particles.

This leads to the question of how large can the KK scale be and
still manage to produce enough signal. We observe that at
$M_{KKglu_1}=1300$ GeV, larger 5D Yukawa couplings ($Y_{5d} \geq 3$)
are required in order to have 
enough signal production. The branching fraction into $hZ$ starts to
be problematic, and actually becomes worse for heavier KK
scales. The signal production requires that the value of
$|c_{u_3}|$ be located around the point where the top Yukawa couplings of
$\Pi_0$ are suppressed, in this case around $|c_{u_3}|\sim 0.3$.

We conclude therefore that in order to explain the $750$ GeV diphoton,
the KK scale must be $M_{KKglu_1} \leq 1300$ GeV, with 5D
Yukawa couplings $Y_{5D}\lsim 3$. For those values, the signal is quite
generic (i.e. small, but typical, $c_{u_3}$ values are required), as
long as the $\Pi_0$ mass is set to $750$ GeV with appropriate 
boundary kinetic terms.

\section{Discussion}
\label{sec:discussion}

We performed an analysis of the scalar sector of warped space models
to investigate whether the {\it minimal} model can accommodate a
resonance at 750 GeV with the properties observed in the diphoton
resonance at the CMS and ATLAS.  We show that in the simplest extra
dimensional  extension of the SM, that is with a 5D Higgs doublet
living in the bulk, the lowest pseudoscalar KK excitation can be
responsible for the signal observed at 750 GeV. We emphasize that,
unlike other explanations relying on scalar fields in warped models,
ours does not introduce any new fields or representations, but relies
on Higgs brane kinetic terms to lower the KK mass of the lightest CP
odd Higgs resonance. This makes the model extremely constrained, with the only new
parameter being the IR brane kinetic coefficient $d_{1}$.  The lightest
CP odd excitation, a mixture of the 5D Higgs field and $Z_5$, does not
decay at tree level into $WW$ or $ZZ$, and, over a range of the parameter space, can
have suppressed couplings to the top quark, thus a small decay width
into $t {\bar t}$. The production through gluon fusion can be
loop-enhanced through the effects of the usual KK fermion modes, and so can
the diphoton decay. The coupling to $Zh$ is also suppressed although
starts increasing dangerously for KK masses above $1500$ GeV.

We also showed that in $AdS_5$ spaces (RS-type models) with a KK scale
as low as $1500$ GeV, the  lowest CP-odd scalar cannot have a mass at
750 GeV, but if the metric differs slightly from the $AdS_5$ metric (generalized warped metric
spaces), the Higgs brane kinetic terms can produce a viable CP-odd scalar of
mass 750 GeV.

Within these modified metric scenarios, and for KK mass scales at around
1 TeV (consistent with precision electroweak bounds) this CP odd resonance
obeys the (current) experimental constraints.  We analyzed 
its production and decay for several values of the lowest KK gluon
mass  ($M_{KKglu_1}$), and show that consistency with the data requires
$M_{KKglu_1} \lsim 1300$ GeV.  

Our analysis is quite general, and should be valid even in the absence
of a  signal at 750 GeV. The general conclusion to be taken from our
analysis here is that warped space models, without any new particles,
can explain a (relatively) light diphoton resonance at the LHC.
Should a diphoton excess be found at higher values, even the RS model
might accommodate such a state without the need modify the
metric.

Among the general features of the light CP-odd scalar resonance,
resulting from the 5D Higgs doublet are the fact that its coupling to top pairs
can be suppressed for appropriate top bulk mass parameters. Also its
coupling to $Zh$ is generically suppressed due to the boundary
conditions of the CP odd state. 
In addition, the model predicts that the spectrum for the CP odd and the charged scalars is
essentially the same since their differential equations and boundary
conditions are almost identical. This means that the lightest charged
Higgs boson is expected to have a mass very close 
to the pseudoscalar mass, so about 750 GeV, in the scenario in which
the latter is the diphoton resonance.

If the diphoton resonance ends up being a real particle, and not a statistical
fluctuation, the prediction from this scenario is that $ZZ$ and $WW$
should not be seen, whereas top pair production, dijet production and
$Zh$ signal should be around the corner.
Moreover a search for the charged scalars at around $750$ GeV might
prove useful in order to disentangle the different models explaining
the resonance.

\section{Acknowledgments}

MF  and MT thank NSERC and FRQNT for partial financial support under grant
numbers SAP105354  and PRCC-191578.

%%%%%%%%%%%%%%%%%%%%%%%%%%%%%%%%%%%%%%%%%%%%%%%%%%%%%%%%%%%%%%%%%%%%%%%%%%%%

\section{Appendix}

In this section we consider the effect of brane localized kinetic
terms associated with the 5D Higgs doublet and also with the gauge
bosons. For simplicity, let's consider a 5D toy model with a Higgs
scalar $H(x,y)$ charged under a local $U(1)$, defined by the following action:
\bea
S&=&\int d^4xdy \sqrt{g}\ \left(-\frac{1}{4} F^2_{MN} + |D^M H|^2 - V(H) \right) \\
&&\ \ \ + \sum_i \int d^4xdy \sqrt{g}\ \delta(y-y_i) \left(\frac{1}{4} r_i F^2_{MN} + d_i  |D^M H|^2 - \lambda_i(H))\right)   
\label{5Daction}
\eea
where $F_{MN} = \partial _M A_N - \partial_N A_M$ and for simplicity
we set gauge coupling constant to unity in the appropriate mass dimensions.
The background spacetime metric is assumed to take the form 
\bea
ds^2=e^{-2\sigma(y)}\eta_{\mu\nu} dx^\mu dx^\nu - dy^2
\eea
where $\sigma(y)$ is the warp factor.

We are interested in studying the effective 4D perturbative spectrum of the 5D Higgs
field and the 5D gauge boson, around a nontrivial Higgs vacuum profile solution $<H>=v(y)$.
\bea
H(x,y)= \frac{1}{\sqrt{2}}\left(v(y) + h(x,y)\right) e^{i \pi(x,y)}
\eea
In particular we are interested in the CP-odd Higgs perturbations
$\pi(x,y)$, whose equations of motion are coupled with the gauge boson
perturbations. The equations read
\bea
&& (1+r_i \delta_i) \partial_\mu \partial^\mu A_\mu - \left( (1+r_i
\delta_i)e^{-2\sigma} A'_\mu\right)' + (1+d_i \delta_i) M_A^2 A_\mu
+\non\\
&&\hspace{3cm}\partial_\mu \left((1+d_i \delta_i) M_A^2 \pi - (1+r_i \delta_i)\partial^\nu A_\nu -
\left((1+r_i \delta_i)e^{-2\sigma} A_5\right)'\right) = 0\\
&&(1+r_i \delta_i) \partial_\mu \partial^\mu A_5- (1+r_i \delta_i) \partial^\nu A'_\nu+ (1+d_i \delta_i)M_A^2
(\pi'-A_5)=0\\
&&(1+d_i \delta_i) \partial_\mu \partial^\mu \pi - (1+d_i \delta_i) \partial^\nu A_\nu +
M_A^{-2}\left((1+d_i \delta_i) M_A^2 e^{-2\sigma} (\pi'-A_5)\right)'= 0
\eea
where $M_A= v(y) e^{-\sigma}$ and where $\displaystyle d_i \delta_i \equiv \sum_i
d_i \delta(y-y_i) $ and $\displaystyle r_i \delta_i \equiv \sum_i
r_i \delta(y-y_i) $.

We fix partially the 5D gauge by imposing
\bea
(1+d_i \delta_i) M_A^2 \pi - (1+r_i \delta_i)\partial^\nu A_\nu -
\left((1+r_i \delta_i)e^{-2\sigma} A_5\right)' = 0. \label{gaugefix}
\eea
The previous gauge fixing equation reads in the bulk:
\bea
M_A^2 \pi - \partial^\nu A_\nu - \left(e^{-2\sigma} A_5\right)' =
0. \label{bulkgaugefix}
\eea
Note that if we evaluate the bulk constrain Eq.~(\ref{bulkgaugefix}) at $y=y_1-\epsilon$
(i.e. right before the IR brane), we obtain:
\bea
 \partial^\nu A_\nu \Big|_{y_1-\epsilon} = M_A^2 \pi - \left(e^{-2\sigma} A_5\right)'\Big|_{y_1-\epsilon}.
 \label{epsiloncondition}
 \eea
On the other hand, the effect of the delta functions in
Eq.(\ref{gaugefix}) is to produce a discontinuity in the 5D field
$A_5$ at the brane location as
\bea
d_1 M_A^2 \pi -r_1 \partial^\nu A_\nu\Big|_{y_1-\epsilon} = \left[e^{-2\sigma} A_5\right]_{y_1-\epsilon}^{y_1},
\eea
and similarly for the UV brane.
We can thus multiply (\ref{epsiloncondition}) by $r_1$ and use it in
the previous equation, and find the necessary boundary condition
between $\pi$ and $A_5$, which ensures that $A_\mu$ is completely decoupled, even on the brane.
We find
\bea
(d_1-r_1) M^2_A \pi + r_1 \left(e^{2\sigma} A_5\right)'
\left. \vphantom{\int^A_A} \right|_{y_1-\epsilon} =-
\left.e^{-2\sigma} A_5\right|_{y_1-\epsilon} \label{boundaryconditions}
\eea
where we have taken $A_5$ to vanish exactly on the brane, but it jumps
right before the boundary.

Inserting the gauge choice in the coupled equations of motion, one
manages  to decouple the gauge modes $A_\mu$ (in both the
bulk and the branes) with a bulk equation
\bea
\partial_\mu \partial^\mu A_\mu - \left(e^{-2\sigma} A'_\mu\right)' +  M_{A}^2 A_\mu =0
\eea
and jump condition on $A'_\mu$
\bea
r_1  \partial_\mu \partial^\mu A_\mu +  d_1 M_{A}^2 A_\mu
\Big|_{y_1-\epsilon}=- \left. e^{-2\sigma}
  A'_\mu\right|_{y_1-\epsilon} 
    \eea
where  $A'_\mu$ again vanishes exactly on the brane, but has a jump
right before it. We separate variables
\bea
A_\mu(x,y)= V^{4d}_\mu(x) V_y(y)
\eea
and find the separated equations for the gauge boson tower become
  \bea
&& \partial_\mu \partial^\mu V^{4d}_\mu(x)+m^2_A V^{4d}_\mu(x)=0\\
&& \left( e^{-2\sigma} V'_y\right)' + ( m^2_A - M_A^2 )  V_y =0
  \eea
  with jump conditions on $V'_y$
  \bea
(d_i M_{A}^2- r_i m^2_A)  V_y \Big|_{y_1-\epsilon}= -\left. e^{-2\sigma} V'_y\right|_{y_1-\epsilon} 
  \eea
where the 4D effective mass $m^2_A$ is the constant of separation of variables.

The remaining equations are, in the bulk,
\bea
&&  \partial_\mu \partial^\mu A_5+ M_A^2 (\pi'-A_5) - \left(M_A^2 \pi\right)' +\left( \left(e^{-2\sigma} A_5\right)' \right)' =0\\
&& \partial_\mu \partial^\mu \pi + M_A^{-2} \left( M_A^2 e^{-2\sigma}
(\pi'-A_5) \right)' -  M_A^2 \pi + \left(e^{-2\sigma} A_5\right)'= 0\label{a5pimixedequation}
\eea
and the fields must verify the boundary conditions of Eq.(\ref{boundaryconditions}).

We now perform a mixed separation of variables:
\bea
A_5(x,y)&=& G(x) g(y) + \pi_x(x) \eta(y) \label{expansion1}\\
\pi(x,y)&=& G(x) h(y) + \pi_x(x) \xi(y)\label{expansion2}
\eea
which is to say that both $A_5(x,y)$ and $\pi(x,y)$ contain each some Goldstone and CP-odd degree
of freedom. The profiles $g(y)$, $\eta(y)$,  $h(y)$ and  $\xi(y)$
quantify how much of each they contain. 
Of course the functions $g$ and $h$ are related to each other, as well as $\eta$ and
$\xi$. The relationships are such that $G(x)$ and $\pi_x(x)$ decouple. 
With the choice
\bea
h(y)&=& \frac{K(y)}{m^2_G}\\
g(y)&=&\frac{K'(y)}{m^2_G}\\
\eta(y)&=&\frac{e^{2\sigma}}{m^2_\pi}X(y)\\
\xi(y)&=&\frac{1}{m^2_\pi M^2_A}X'(y)
\eea
and using the mixed separation of variables in (\ref{expansion1}) and
(\ref{expansion2}), the mixed equations of
motion in (\ref{a5pimixedequation}) decouple
and we obtain
\bea
&& \frac{e^{2\sigma}}{M_A^2} X(y)  \partial_\mu \partial^\mu \pi_x(x) +\pi_x(x) e^{2\sigma} X(y)
- \pi_x(x) \left[M_A^{-2} X'(y)\right]'= 0\\
&&K(y) \partial_\mu\partial^\mu G(x)+ G(x) \left[ \left(K'e^{-2\sigma}\right)' + M_A^2  K(y)\right] = 0 .
\eea
Once separated, we obtain, for the CP odd physical scalars
\bea
&&\partial_\mu \partial^\mu \pi_x(x) + m_\pi^2 \pi_x(x) =0\\
&&\left(M_A^{-2} X'\right)'+ e^{2\sigma} \left(\frac{m_\pi^2}{M_A^2}-1\right) X=0 ,
\eea
with boundary conditions
\bea
d_i X' = -X
\eea
and for the Goldstone modes,
\bea
&&\partial_\mu\partial^\mu G(x)+ m_G^2 G(x)=0\\
&&\left(K'e^{-2\sigma}\right)'+ (m_G^2-M_A^2)  K=0, 
\eea
with boundary conditions
\bea
(d_i M_A^2-r_i m^2_G ) K = -e^{-2\sigma} K' .
\eea
Note that both the equations and boundary conditions for the
Goldstone bosons are identical to the ones for the gauge boson tower, as 
should be, so that they can then be gauged away level by level with the
remaining gauge fixing freedom.

\end{document}